\documentclass{aa}  

\usepackage{graphicx,xcolor,natbib,hyperref}
\usepackage{txfonts,textcomp}
\usepackage{multirow}
\usepackage{soul}
\usepackage{array}
\usepackage{lscape}
\usepackage[normalem]{ulem}
\defcitealias{Sharma24}{Paper~I}

\begin{document} 

   \title{Chemical composition of planetary hosts}
   \subtitle{II. Abundances of neutron-capture elements
   \thanks{This work is based on observations collected with the VUES spectrograph installed on a 1.65~m telescope set up at the Mol\.{e}tai Astronomical Observatory, Institute of Theoretical Physics and Astronomy, Vilnius University, Lithuania (A facility that belongs to the Europlanet Telescope Network \url{https://www.europlanet-society.org/europlanet-2024-ri/europlanet-telescope-network/}).}
   }
   
   \author{A. Sharma
           \inst{1},          
           E. Stonkut\.{e}
           \inst{1},
           A. Drazdauskas
           \inst{1},
           R. Minkevi\v{c}i\={u}t\.{e}
           \inst{1},
           \v{S}. Mikolaitis
           \inst{1},
           G. Tautvai\v{s}ien\.{e}
           \inst{1},
           \and
           U. Jonauskait\.{e}
           \inst{1}
          }

   \institute{Vilnius University, Faculty of Physics, Institute of Theoretical Physics and Astronomy, 
              Sauletekio av. 3, 10257, Vilnius, Lithuania\\
              \email{ashutosh.sharma@ff.vu.lt}
             }

   \date{}

   \titlerunning{Chemical composition of planetary hosts: Abundances of neutron-capture elements}
   \authorrunning{Sharma et al.}

 
  \abstract
   {}
   {This study seeks to determine abundances of neutron-capture elements (Sr, Y, Zr, Ba, La, Ce, Nd, Pr, and Eu) in a large and homogeneous sample of F, G, and K-type planet-host stars (PHSs) located in the northern hemisphere. The sample includes 160 stars, 32 of which are in multi-planetary systems. These stars host a total of 175 high-mass planets and 47 Neptunian and Super-Earth planets. We investigated potential correlations between stellar chemical compositions and the presence of orbiting planets.}
   {Spectra were obtained using the 1.65-metre telescope at the Mol{\. e}tai Astronomical Observatory and a fibre-fed high-resolution spectrograph covering the entire visible wavelength range (4000--8500~$\AA$). The abundances of neutron-capture elements were determined by differential line-by-line spectrum synthesis using the TURBOSPECTRUM code with the MARCS stellar model atmospheres.}
   {We analysed neutron-capture elements relative to iron ([El/Fe]) and found that the abundances of the majority of chemical elements in exoplanet host stars align with the Galactic chemical evolution. However, [Zr/Fe], [La/Fe], and [Ce/Fe] are overabundant in stars with planets compared to reference stars at a given [Fe/H]. When examining [El/Fe] against planet mass, most elements show positive correlations with higher mass planets, excluding strontium, yttrium, and barium, which exhibit insignificant correlations across all sub-samples. The $\Delta$[El/H] versus $T_{\text{cond}}$ slope distribution shows a positive skewness for planet-hosting stars, suggesting an enrichment of refractory elements compared to analogues. While $\Delta$[El/H]–$T_{\text{cond}}$ slopes and stellar and planetary parameters do not show strong correlations, trends suggest that older dwarf stars with multiple planets have smaller or even negative $\Delta$[El/H]–$T_{\text{cond}}$ slopes compared to younger dwarf stars, which show larger positive slopes. Our results also show that multi-planetary systems are more common around metal-rich stars.}
   {}


   \maketitle

\section{Introduction}
Since the discovery of the first exoplanets \citep{Wolszczan92, Mayor95}, more than 5,900 have been confirmed\footnote{\url{https://exoplanetarchive.ipac.caltech.edu/}} around a diverse spectrum of host stars, and the search continues. Detailed studies of planet-host stars (PHSs) are crucial for upcoming research (e.g. PLATO:~\citealt{Rauer24}; Ariel:~\citealt{Tinetti22}; E-ELT:~\citealt{deZeeuw14}) and enhancing the characterisation of exoplanets, leading to a better understanding of their formation and evolution. The characteristics of exoplanets are closely linked to both their host stars and the environments in which those stars originated and exist. Consequently, we are conducting a follow-up programme of PHSs using the high-resolution Vilnius University Echelle Spectrograph \citep{Jurgenson16} at the Mol\.{e}tai Observatory's 1.65 m telescope. In \citeauthor{Sharma24} (2024; hereinafter called Paper~I), we determined the main atmospheric parameters and chemical abundances of C, N, O, Mg, and Si, along with C/O, N/O, and Mg/Si abundance ratios in stars with planets. In this work, we complement our published results with a homogeneous abundance analysis of neutron-capture (\textit{n}-capture) elements of diverse $s/r$--process inputs in their production (Sr, Y, Zr, Ba, La, Ce, Nd, Pr, and Eu). 

Most of the isotopes beyond iron are produced by rapid or slow \textit{n}-capture processes (hereafter \textit{r} and \textit{s} processes). These processes differ in neutron density, which affects the likelihood of a neutron being captured before radioactive beta decay occurs \citep{Burbidge57}. The majority of \textit{n}-capture elements are formed through a combination of \textit{r} and \textit{s} processes, with only a few cases where one process predominantly contributes to their production. The \textit{s} process can be divided into sub-processes: the weak \textit{s} process, which primarily produces the so called first peak of \textit{s}-process elements (Sr, Y, Zr); and the main \textit{s} process, responsible for generating the second peak of \textit{s}-process elements (Ba, La, Pr, Ce, Nd). Additionally, Eu is considered a pure \textit{r}-process element \citep{Travaglio99, Bisterzo14, Prantzos20}.

The \textit{s} process takes place during He-core and convective-shell C burning of massive stars and in the inter-shell region of low- and intermediate-mass asymptotic giant branch (AGB) stars \citep{Karakas14, Cseh22}. The \textit{r} process is not as well understood than the \textit{s} process, and there is still no consensus as to where the \textit{r} process takes place. Supernova explosions, neutron-star mergers, and other cosmic events have each been proposed as sites hosting the \textit{r} process (\citealt{Cowan21} and references therein).

This work aims to advance our previous research by focusing on heavier elements in PHSs, which have not been investigated as thoroughly in the literature as light or alpha elements. \cite{Bond08} conducted one of the first studies of 28 PHSs to derive abundances of five \textit{r}- and \textit{s}-process elements (Y, Zr, Ba, Nd, and Eu). They found that host stars are enriched in all studied elements compared to non-host stars, with mean differences ranging from 0.06 to 0.11~dex. The study concluded that the chemical anomalies in planetary host stars result from the normal Galactic chemical evolution processes. In contrast, \cite{daSilva15} analysed 47 stars with planets and found the result for barium intriguing and warranting further investigation. Similarly, \cite{Mishenina16}, which examined several heavy elements in 14 PHSs, reported an underabundance of Ba in exoplanet-hosting stars. \citet{Delgado18} analysed a sample of 151 stars with planets and derived a deficiency of Ba in low-mass planet hosts. They concluded that to make a fair comparison with stars that do not have detected planets, it is essential to understand and characterise the stellar populations to which planet hosts belong.

Heavy-element abundances are essential for constraining models of Galactic chemical evolution and understanding the yields of both massive and low-mass stars. It is important to decipher how we can relate the properties of stars and planets to planet formation. For example, we know that giant planets occur more frequently around stars with higher metallicities (usually taken as iron abundance) \citep{Fischer05, Ghezzi10}. The planet signatures appear as correlations between elemental abundance differences and the dust condensation temperature \citep[see e.g.][]{Melendez09}. This naturally leads us to ask if \textit{n}-capture elements play a role in planetary formation. In this paper, we present the results of \textit{n}-capture element analysis in stars with planets and address several questions about the star-planet connection that are still under discussion. These include the influence of stellar chemical composition and the potential role of the planetary mass \citep[e.g.][]{Mishenina16, Delgado18, Swastik22, daSilva24} and the relationship between element condensation temperatures and their elemental abundances \citep[e.g.][]{Melendez09, Adibekyan16,Yun24}.

This work is organised as follows. In Sect.~\ref{sec:data}, we describe the stellar sample and present the methodology for determining the chemical abundances of \textit{n}-capture elements. In Sect.~\ref{sec:results}, we present the results of our analysis along with a detailed discussion. Finally, in Sect.~\ref{sec:conclusions}, we highlight the main conclusions drawn from this study.

\section{Stellar sample and analysis}
\label{sec:data}

\subsection{Stellar sample}
Our sample comprises 160 bright stars (V$\leq$8.5 mag) selected from the Transiting Exoplanet Survey Satellite (TESS) catalogue \citep{Rinehart15} and observed using the 1.65-metre Ritchey-Chrétien telescope located at the Mol\.{e}tai Astronomical Observatory in Lithuania. High-resolution spectra were acquired with the Vilnius University Echelle Spectrograph (VUES; \citeauthor{Jurgenson16}\citeyear{Jurgenson16}). Detailed information on the observations of the 149 targets can be found in \citetalias{Sharma24}. We reviewed the  Mol\.{e}tai Astronomical Observatory archive for spectra of previously observed stars \citep{Stonkute20, Tautvaisiene20, Tautvaisiene22} and identified 11 additional stars with recently confirmed planets, expanding our PHSs sample to a total of 160 stars. This final sample consists of 86 main-sequence stars (hereafter referred to as dwarf stars) along with 74 stars that are at their evolved stages (hereafter referred to as giant stars). Additionally, we compiled a comparison sample comprising 491 stars (222 dwarfs and 269 giants) with previously determined chemical abundances from \cite{Tautvaisiene21}, which currently have no detected planetary companions. 

\subsection{Stellar atmospheric parameters and abundances}
\label{subsec:parameters}
The main atmospheric parameters required for the abundance analysis were adopted from our previous study of 149 PHSs \citepalias{Sharma24}, while for the additional 11 stars with confirmed planets, these parameters were sourced from \citet{Stonkute20} and \citet{Tautvaisiene20, Tautvaisiene22}. The parameters were derived from high-resolution spectroscopic observations using a well-established equivalent-width method. The equivalent widths of neutral and ionised iron lines (hereafter referred to as \ion{Fe}{i} and \ion{Fe}{ii}, respectively) were measured using the DAOSPEC software \citep{Stetson08}. Additionally, we used the MOOG program \citep[][version 2014]{Sneden73} to calculate the iron abundances that were subsequently used to infer the atmospheric parameters of the stars.

   \begin{figure}
    \centering
    \includegraphics[width=0.9\hsize]{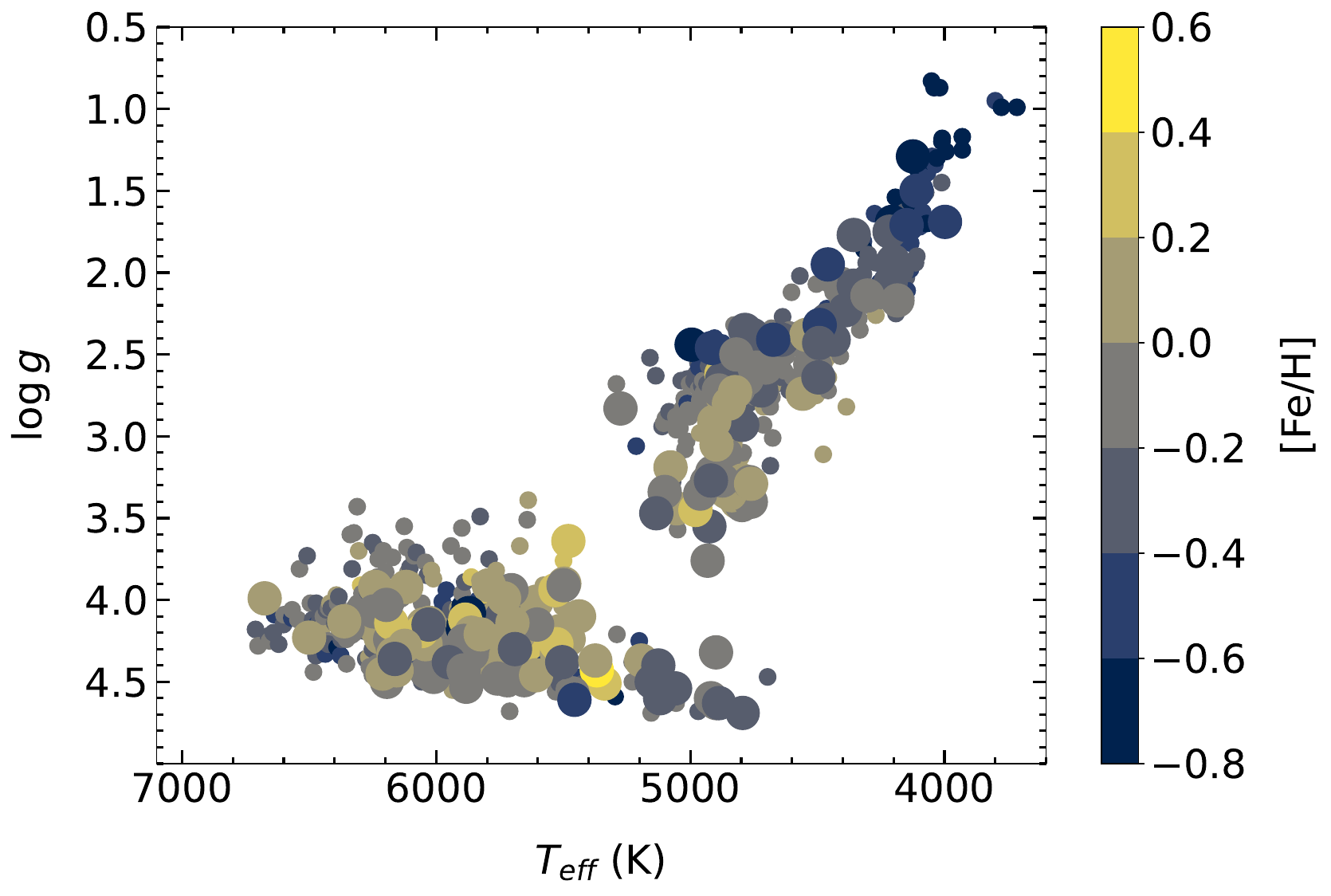}
    \caption{Effective temperature ($T_{\rm eff}$) versus surface gravity (${\rm log}~g$) diagram for analysed planet host stars, colour-coded by metallicity, [Fe/H]. The comparison stars marked with smaller symbols are from \citet{Tautvaisiene21}.}
    \label{fig:Teffvslogg}
   \end{figure}

   \begin{figure*}
    \centering
    \includegraphics[width=0.95\hsize]{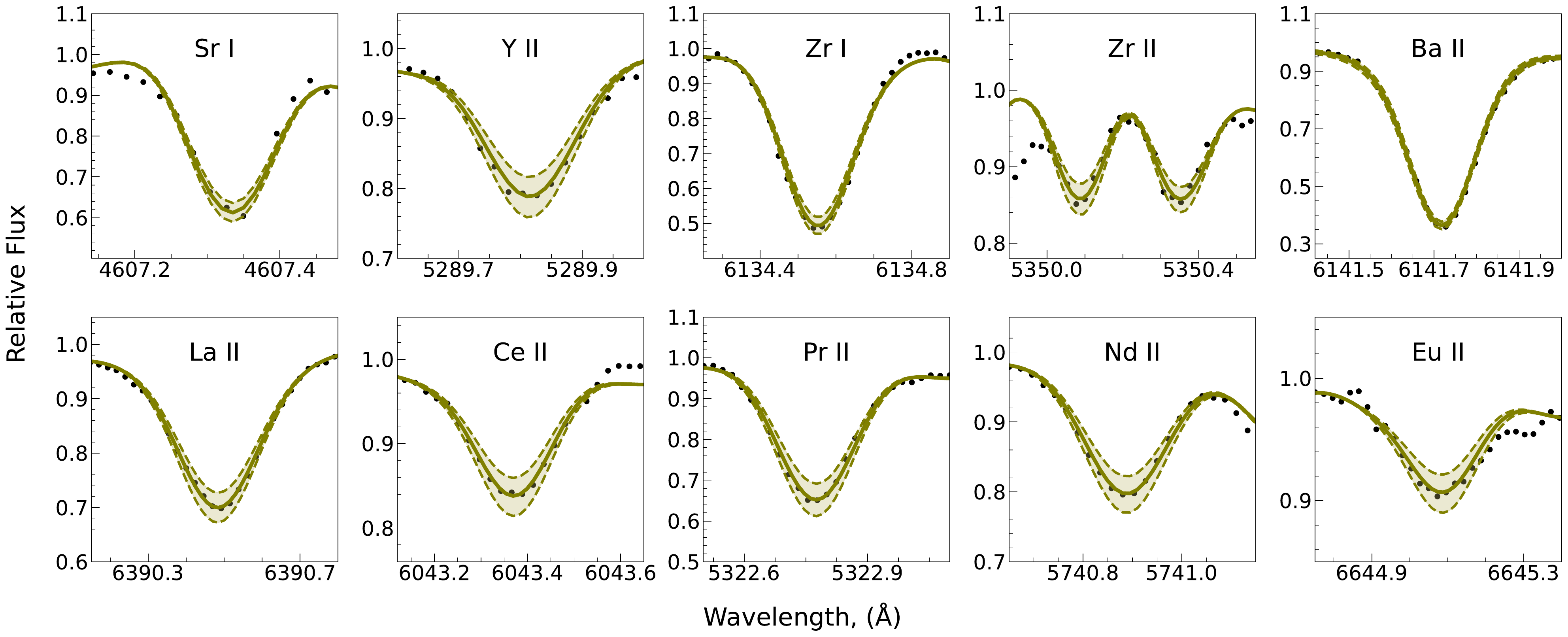}
    \caption{Examples of synthetic spectrum fits for neutron-capture elements in our stars with planets. The observed spectra (black dots) are compared with the best-fit synthetic spectra (solid olive lines) to determine elemental abundances. Dashed lines represent the variations of ±0.10 dex from the optimally fitted abundance.}
    \label{fig:spectrumfits}
   \end{figure*}

In summary, the effective temperature, $T_{\rm eff}$, was determined by minimising the slope of iron abundances derived from \ion{Fe}{i} lines as a function of the excitation potential. The surface gravity, ${\rm log}~g$, was determined by ensuring the ionisation equilibrium, such that the Fe\,{\sc i} and Fe\,{\sc ii} lines give the same iron abundances. Finally, the microturbulence velocity, $v_{\rm t}$, was determined by requiring that the iron abundances from \ion{Fe}{i} lines remain independent of the equivalent width. Figure~\ref{fig:Teffvslogg} presents the investigated stars in an effective temperature versus surface gravity diagram, with the comparison stars from \citet{Tautvaisiene21} represented by smaller symbols. The analysed sample is approximately evenly distributed between evolved and main-sequence stars.

The chemical abundances of heavy elements in our stellar sample were determined by analysing the stellar spectra using the same differential model atmosphere technique as presented in \citetalias{Sharma24}. Moreover, heavy elements often comprise several isotopes, and their spectral lines can be affected by hyperfine splitting. To accurately determine the abundances for these elements, it is essential to use spectral line synthesis, which accounts for wavelength shifts of various isotopes, hyperfine components, and other blending features.

The spectral synthesis was performed for all investigated chemical elements using a differential approach relative to the Sun, employing a line-by-line comparison of the high-resolution observed spectra to the modelled spectra. We used the Turbospectrum code \citep{Alvarez98} to generate synthetic stellar spectra using the main atmospheric parameters of our host stars. The calculations were based on a grid of plane-parallel, one-dimensional (1-D), hydrostatic model atmospheres under the assumption of local thermodynamic equilibrium (LTE) obtained from the MARCS stellar model atmosphere library \citep{Gustafsson08}. We used version 5 of the $Gaia$-ESO Survey line list \citep{Heiter21}, selecting lines marked 'Yes' or 'Undecided'. These flags indicate that the line is either 'relatively unblended' or 'potentially useful' in some stars. Spectral lines for each element were selected on the basis of their strength and minimal contamination from neighbouring features.

After finalising the line list, we calibrated it using the solar spectrum from \cite{Kurucz05} with the solar chemical element abundances from \cite{Grevesse07}. We tested the calibrated line list on the solar spectra obtained at Mol\.{e}tai Astronomical Observatory using the same instrumental setup as that used for the sample stars. The derived abundances remained within the uncertainty limits of individual chemical elements, which do not exceed 0.05~dex. In Fig.~\ref{fig:spectrumfits}, we present examples of synthetic spectrum fits for \textit{n}-capture elements in PHSs from our sample. The specific spectral lines used for the abundance analysis of each investigated chemical element are detailed below.

We selected two strontium (Sr\,{\sc i}) spectral lines, one at 4607~{\AA} and the other at 7070~{\AA}. 
The abundances of yttrium were derived from up to seven Y\,{\sc ii} lines at wavelengths 4883, 4900, 4982, 5087, 5200, 5289, and 5402~$\AA$.
Zirconium abundances were measured using both neutral (Zr\,{\sc i}) and ionised (Zr\,{\sc ii}) lines. The selected Zr\,{\sc i} lines were at 6127 and 6134~{\AA}, and the Zr\,{\sc ii} lines were at 5350.1 and 5350.3~{\AA}.
Barium abundances were determined using the Ba\,{\sc ii} lines at 5853, 6141, and 6496~{\AA}. For all Ba\,{\sc ii} lines, the log\,$gf,$  the HFS, and IS values were taken from \citet{Davidson92}. The strong barium lines include a background line list with Ba\,{\sc ii} data from \citet{Miles69}. 
Lanthanum abundances were derived from up to five La\,{\sc ii} spectral lines at wavelengths 4748, 4804, 5123, 5303, and 6390~{\AA}.  The HFS and IS values for the La\,{\sc ii} 5123, 5303, and 6390~{\AA}\, lines were also taken from \citet{Lawler01La}. 
Up to four cerium (Ce\,{\sc ii}) lines at 5274, 5472, 5512, and 6043~{\AA} were used to determine Ce abundances. 
Praseodymium abundances were determined using two Pr\,{\sc ii} lines at 5259 and 5322~{\AA}. The HFS and IS values for all Pr\,{\sc ii} lines were taken from \citet{Sneden09}.
Neodymium abundances were derived from five Nd\,{\sc ii} lines at 5092, 5255, 5276, 5357, and 5740~{\AA}. The HFS and IS values were taken for the Nd\,{\sc ii} 5092 and 5740~{\AA}\, lines from \citet{DenHartog03} and for the 5276~{\AA}\, line from \citet{Meggers75}. 
Europium abundances were derived using the Eu\,{\sc ii} line at 6645~{\AA}. The HFS and IS values for the Eu\,{\sc ii} line were obtained from \citet{Lawler01Eu}.

\subsection{Non-local thermodynamic equilibrium effects}
Our abundance determinations are based on 1-D LTE modelling. However, we also considered recent studies that investigated the effects of non-local thermodynamic equilibrium (NLTE) in abundance determinations. While many previous works have focused on NLTE corrections for metal-poor halo and thick-disc stars, where these effects are most pronounced, our sample consists of PHSs with an average metallicity of <[Fe/H]>=~$-0.09$. The most metal-poor star in our dataset has a metallicity of [Fe/H]=~$-0.76$.

Using the 2020 version of the Turbospectrum code, along with the calculate\_nlte\_correction\_line subroutine from the TSFitPy wrapper \citep{Gerber23}, we computed NLTE corrections for the elements: Sr \citep{Bergemann12a, Gerber23}, Y \citep{Storm23,Storm24}, Ba \citep{Gallagher20}, Eu \citep{Storm24}, and Mg \citep{Bergemann17}.

The strontium spectral lines exhibit varying sensitivity to NLTE effects. In particular, the line at 4607~{\AA} is significantly more affected than the one at 7070~{\AA}. Our NLTE calculations for the 4607~{\AA} line yielded corrections ranging from +0.02 to +0.12 dex for dwarf stars, with a median correction of +0.07 dex. In contrast, giant stars showed stronger NLTE effects, with corrections ranging from +0.06 to +0.30 dex and a median correction of +0.16 dex. These corrections tend to increase with decreasing metallicity. On the other hand, the NLTE analysis of the Sr\,{\sc i} 7070~{\AA} line revealed negligible corrections across our sample, with values effectively centered around zero (median correction $\approx$ 0.00 dex), and extreme values remaining within 0.00001 dex.

The analysed yttrium spectral lines can be grouped into two categories based on the magnitude of their NLTE corrections. The first group --comprising the lines at 4883, 4900, 5087, and 5200~{\AA--} exhibits more pronounced NLTE effects, particularly in giant stars. For dwarf stars, the corrections for these lines range from –0.04 to +0.07~dex, while for giants the corrections span from –0.03 to +0.14 dex. Despite this range, the average corrections are close to zero: +0.01 dex for dwarfs and +0.04 dex for giants. In contrast, the second group of lines shows minimal sensitivity to NLTE effects. For these lines, the corrections range from –0.00 to +0.03 dex in dwarfs and from –0.00 to +0.04 dex in giants. The mean corrections are +0.01 dex and +0.02 dex for dwarfs and giants, respectively.

Barium lines are often challenging to fit due to their considerable strength and large equivalent widths, which in turn necessitate more substantial NLTE corrections. These corrections are comparable in both dwarf and giant stars, although their behaviours differ from those observed for Sr and Y, with slightly larger corrections found in dwarfs. For the Ba\,{\sc ii} line at 5853~{\AA}, the NLTE corrections are the smallest, ranging from –0.12 to –0.02 dex in dwarf stars and from –0.17 to –0.02 dex in giants. The other two lines, at 6141~{\AA} and 6496~{\AA}, show very similar behaviour. In dwarfs, the 6141~{\AA} corrections span approximately –0.24 to –0.03 dex, while those for the 6496~{\AA} line range from –0.25 to –0.05 dex; for giants, the ranges are about –0.19 to –0.00 dex for the 6141~{\AA} line and –0.26 to –0.03 dex for the 6496~{\AA} line. On average, the corrections for these lines are roughly –0.13 dex in dwarfs and –0.08 dex in giants.

For europium abundance determination, we used a single Eu,{\sc ii} line at 6645~{\AA}. This line shows some sensitivity to NLTE effects, with similar correction patterns observed in both dwarf and giant stars. The average NLTE correction for dwarfs is –0.05 dex, with values ranging from –0.08 to +0.03 dex. For giants, the corrections span from –0.11 to –0.00 dex, with an average of –0.05 dex.

Although magnesium abundances were determined in our previous work, we recalculated NLTE corrections for all lines used in that study. For our line list and stellar sample, the NLTE effects are negligible. The largest line-by-line corrections range from –0.08 to 0.00 dex. However, when averaged across all lines for individual stars, the corrections fall between –0.03 and 0.00 dex, indicating no significant impact on the final abundance results.

In the study by \cite{Shaltout20}, NLTE effects on the praseodymium abundance were investigated in the solar atmosphere using 14 Pr\,{\sc ii} lines, including the 5259 and 5322~{\AA} lines adopted in our study. Their analysis found a minimal NLTE abundance correction of approximately +0.01 dex, suggesting that under solar-like conditions NLTE effects on Pr\,{\sc ii} lines are negligible. \cite{Mashonkina05} investigated the NLTE ionisation equilibrium of Nd\,{\sc ii} and Nd\,{\sc iii}, focusing primarily on the cool A and Ap stars. Their findings indicated significant NLTE effects in these hotter stellar atmospheres, particularly at effective temperatures between 7\,500 K and 9\,500 K. However, the relevance of these results to cooler FGK-type stars remains uncertain.

\subsection{Determination of uncertainties} 
\label{subsec:errors}

   \begin{table}
    \caption{Effect of uncertainties in atmospheric parameters on derived chemical abundances.} 
    \label{table:uncertainities}      
    \centering
    \renewcommand{\arraystretch}{1.30}
    \begin{tabular}{l c c c c}
    \hline
    \hline
    {\raisebox{-1.5ex}[0cm][0cm]{Elements}}  &  $\Delta T_{\rm eff}$  &  $\Delta {\rm log}~g$  &  $\Delta {\rm [Fe/H]}$  &  $\Delta v_{\rm t}$ \\
    &  $\pm$ 50\,\text{K}  &  $\pm$ 0.20\,\text{dex}  &  $\pm$ 0.09\,\text{dex}  &  $\pm$ 0.25\,km\,s$^{-1}$ \\
    \hline
    \multicolumn{5}{l}{Dwarfs (${\rm log}~g$ > 3.5)} \\
    \hline
    Sr\,{\sc i}       &  $\pm$ 0.06  &  $\mp$ 0.01  &  $\pm$ 0.01  &  $\mp$ 0.08   \\  
    Y\,{\sc ii}       &  $\pm$ 0.01  &  $\pm$ 0.07  &  $\pm$ 0.02  &  $\mp$ 0.12   \\
    Zr\,{\sc i}       &  $\pm$ 0.06  &  $\pm$ 0.01  &  $\mp$ 0.01  &  $\pm$ 0.01   \\      
    Zr\,{\sc ii}      &  $\pm$ 0.01  &  $\pm$ 0.08  &  $\pm$ 0.02  &  $\mp$ 0.01   \\
    Ba\,{\sc ii}      &  $\pm$ 0.02  &  $\pm$ 0.02  &  $\pm$ 0.01  &  $\mp$ 0.14   \\
    La\,{\sc ii}      &  $\pm$ 0.02  &  $\pm$ 0.08  &  $\mp$ 0.05  &  $\mp$ 0.01   \\
    Ce\,{\sc ii}      &  $\pm$ 0.02  &  $\pm$ 0.08  &  $\mp$ 0.03  &  $\mp$ 0.01   \\
    Pr\,{\sc ii}      &  $\pm$ 0.01  &  $\pm$ 0.09  &  $\pm$ 0.02  &  $\pm$ 0.01   \\
    Nd\,{\sc ii}      &  $\pm$ 0.01  &  $\pm$ 0.09  &  $\pm$ 0.03  &  $\pm$ 0.01   \\
    Eu\,{\sc ii}      &  $\pm$ 0.01  &  $\pm$ 0.09  &  $\pm$ 0.02  &  $\pm$ 0.01   \\
    \hline
    \multicolumn{5}{l}{Giants (${\rm log}~g$ $\leq$ 3.5)}\\
    \hline
    Sr\,{\sc i}       &  $\pm$ 0.05  &  $\mp$ 0.01  &  $\mp$ 0.01  &  $\mp$ 0.01   \\  
    Y\,{\sc ii}       &  $\pm$ 0.01  &  $\pm$ 0.08  &  $\pm$ 0.02  &  $\mp$ 0.18   \\
    Zr\,{\sc i}       &  $\pm$ 0.09  &  $\pm$ 0.01  &  $\mp$ 0.01  &  $\mp$ 0.01   \\      
    Zr\,{\sc ii}      &  $\mp$ 0.01  &  $\pm$ 0.09  &  $\pm$ 0.03  &  $\mp$ 0.02   \\
    Ba\,{\sc ii}      &  $\pm$ 0.04  &  $\pm$ 0.04  &  $\mp$ 0.01  &  $\mp$ 0.20   \\
    La\,{\sc ii}      &  $\pm$ 0.01  &  $\pm$ 0.08  &  $\pm$ 0.03  &  $\mp$ 0.02   \\
    Ce\,{\sc ii}      &  $\pm$ 0.01  &  $\pm$ 0.09  &  $\pm$ 0.01  &  $\mp$ 0.04   \\
    Pr\,{\sc ii}      &  $\pm$ 0.01  &  $\pm$ 0.08  &  $\pm$ 0.04  &  $\mp$ 0.01   \\
    Nd\,{\sc ii}      &  $\pm$ 0.01  &  $\pm$ 0.08  &  $\pm$ 0.03  &  $\mp$ 0.03   \\
    Eu\,{\sc ii}      &  $\mp$ 0.01  &  $\pm$ 0.09  &  $\pm$ 0.03  &  $\pm$ 0.01   \\
    \hline
    \end{tabular} 
   \end{table}

   \begin{figure*}
    \centering
    \includegraphics[width=0.7\hsize]{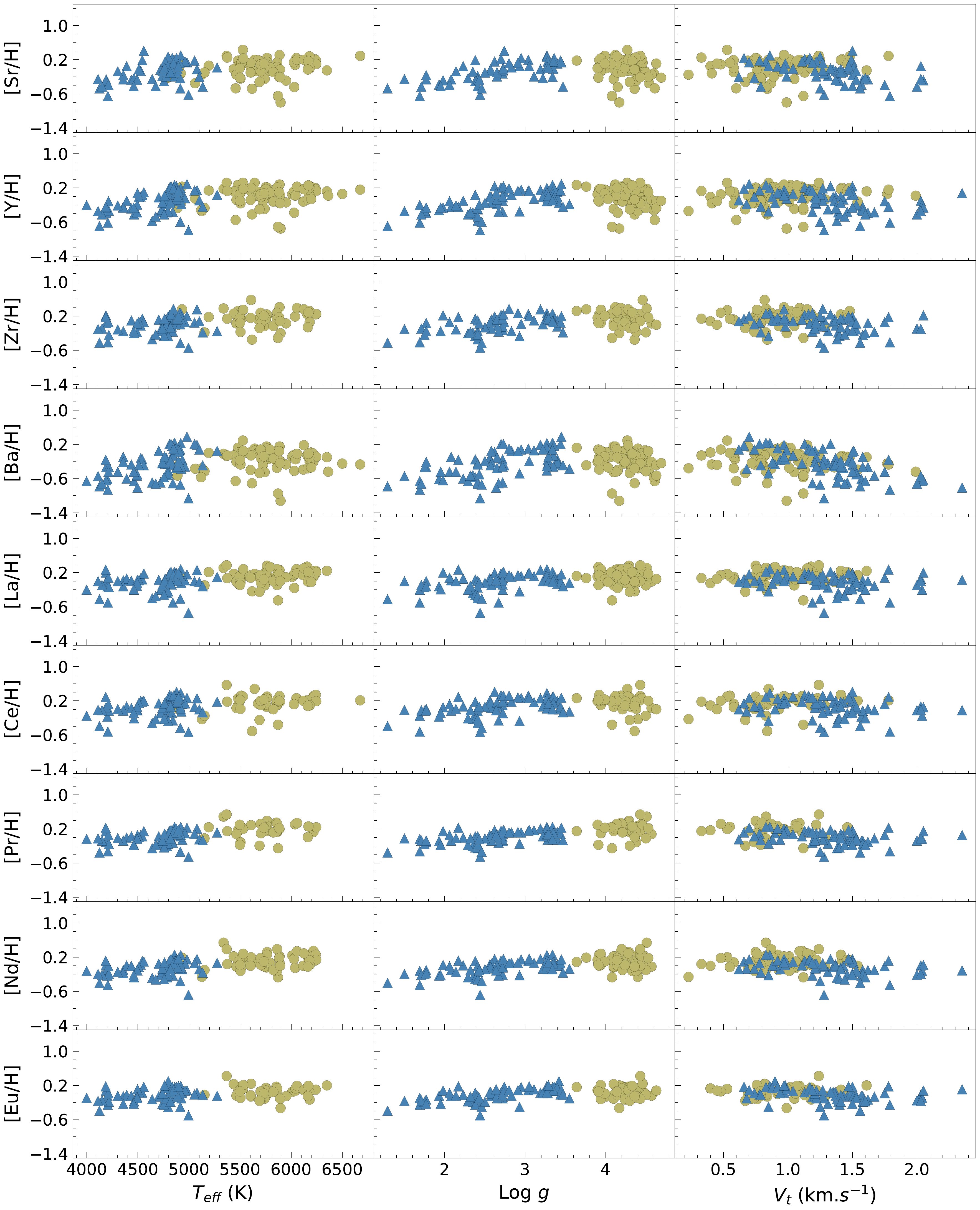}
    \caption{Abundances of neutron-capture elements as a function of atmospheric parameters. Dwarfs and giants are represented by circles and triangles, respectively. The Sr, Y, Ba, and Eu abundances include the non-LTE corrections.}
    \label{fig:A(X)vsparams}
   \end{figure*}

   \begin{figure*}
    \centering
    \includegraphics[width=0.8\hsize]{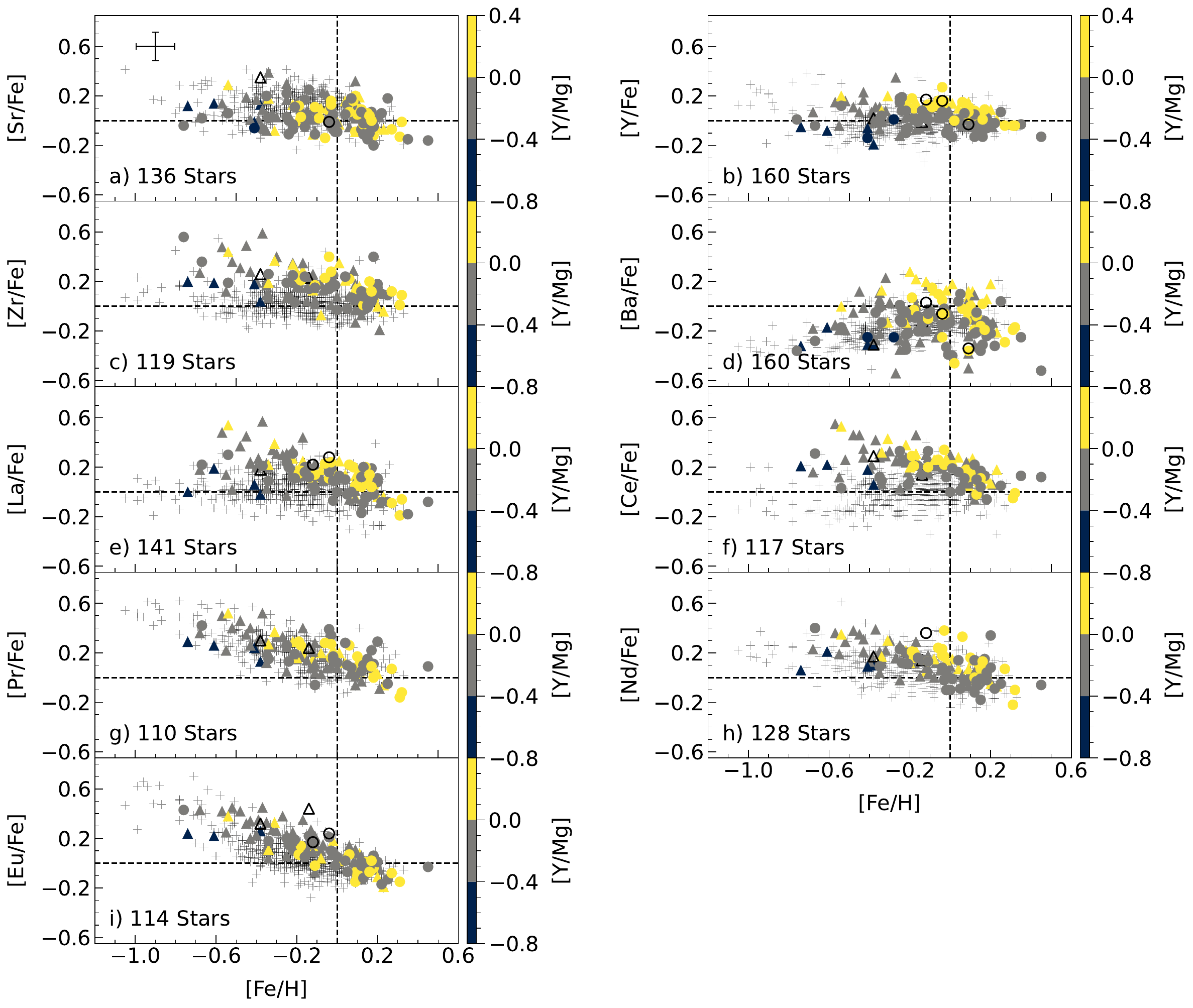}
    \caption{Abundances of neutron-capture elements [El/Fe] as functions of metallicity [Fe/H] for all investigated stars. Dwarf stars are represented by circles, while giant stars (with ${\rm log}~g$ $\leq$ 3.5) are denoted by triangles. The stars are colour-coded according to their [Y/Mg] ratio. Average error bars are displayed in panel \textit{a}. The comparison sample of disc stars is denoted by plus signs and taken from \cite{Tautvaisiene21}.}
    \label{fig:XFevsFeH_YMg}
   \end{figure*}

We carefully addressed potential sources of uncertainties at each stage of the analysis for our stellar sample in \citetalias{Sharma24}. These uncertainties can be systematic and random. Systematic uncertainties, such as those arising from uncertainties in atomic data, were largely minimised through the use of differential analysis relative to the Sun. Random uncertainties, on the other hand, which primarily arise from factors such as local continuum placement and the fitting of individual spectral lines, were carefully addressed in this work.   

Random uncertainties in the elemental abundances were estimated by deriving abundances from multiple spectral features, with the scatter among individual line measurements serving as a measure of the uncertainty for a given element. This methodology was uniformly applied across all analysed elements except europium, for which the abundance was determined using a single atomic line. In this case, the uncertainty was estimated manually during the spectral synthesis process. Additionally, elemental abundances are sensitive to uncertainties in stellar atmospheric parameters. To evaluate the impact of uncertainties that arise from stellar parameters, we calculated the variations in abundance induced by the estimated uncertainty in each atmospheric parameter while keeping all other parameters fixed. The atmospheric parameters, along with their associated uncertainties for our stellar sample, are provided in \citetalias{Sharma24}. The typical uncertainties in our measurements are as follows: the average uncertainty in the effective temperature, $T_{\rm eff}$, is approximately $50\pm15$~K; that in the surface gravity, ${\rm log}~g$, is $0.20\pm0.05$~dex; that in metallicity, [Fe/H], is $0.09\pm0.02$~dex; and that in the microturbulent velocity, $v_{\rm t}$, is $0.25\pm0.08$~km\,s$^{-1}$.

The sensitivity of abundance determinations to variations in atmospheric parameters is presented in Table~\ref{table:uncertainities}. We calculated the variations in abundances caused by uncertainty in each atmospheric parameter while keeping the other parameters constant. The results indicate that the abundances are generally not very sensitive to changes in atmospheric parameters for both dwarfs and giants. Notable exceptions are barium and yttrium abundances, which are particularly sensitive to changes in the microturbulence velocity.

\section{Results and discussion} 
\label{sec:results}
In this section, we present the results of our analysis based on a combined sample of 160 stars, which includes 149 PHSs from \citetalias{Sharma24} and an additional 11 stars from our previous studies. The fundamental stellar parameters, as well as the kinematic properties, ages, and other relevant stellar and planetary characteristics, are provided in \citetalias{Sharma24}. The values of derived abundances for \textit{n}-capture elements are available at the CDS, and the table format is provided in Table~\ref{table:Results}. 

The main atmospheric parameters for our extended sample of PHSs span the following ranges: $T_{\rm eff}$ between 4000 and 6680~K; ${\rm log}~g$ range from 1.3 to 3.5 for giant stars and from 3.6 to 4.7 for dwarf stars; [Fe/H] varies from $-0.76$ to 0.45~dex, with an average of $-0.09\pm0.24$~dex. The results of our previous studies show that most of the PHSs in our sample exhibit thin-disc kinematics, and our analysis of \textit{n}-capture element abundances is consistent with the Galactic chemical evolution \citep[see e.g.][]{Pagel97, Grisoni20, Matteucci21}. 

In Fig.~\ref{fig:A(X)vsparams}, we present the abundance trends of all studied species as a function of stellar parameters. Dwarf and giant stars are represented by circles and triangles, respectively. The Sr, Y, Ba, and Eu abundances include the non-LTE corrections. The results indicate no systematic differences in the derived abundances.

\subsection{Elemental abundance trends relative to metallicity}
\label{subsec:heavy/Fe}
We explored the trends of the derived elemental abundances of \textit{n}-capture elements, [El/Fe], for our sample of PHSs of the F, G, and K type in relation to stellar metallicity, [Fe/H]. NLTE corrections were applied to Sr, Y, Ba, Eu, Mg, and Si for both PHSs and the comparison sample. Figure~\ref{fig:XFevsFeH_YMg} shows [El/Fe] as a function of [Fe/H]. A comparison sample of stars, denoted by plus signs, is also shown, which were observed with the same telescope and taken from \citet{Tautvaisiene21}. Circles represent dwarf stars, while giant stars are depicted by triangles, with colour-coding based on the [Y/Mg] ratios, where a higher [Y/Mg] ratio suggests a younger age, and vice versa. The few stars where Mg abundances could not be determined are represented by empty circles and/or triangles. We colour-coded our data based on the [Y/Mg] ratio since it can be used as a chemical clock to trace stellar ages. 

\subsubsection{\textit{Strontium, yttrium, and zirconium}}
Sr, Y, and Zr are considered light \textit{s}-process-dominated elements, as they are part of the first \textit{s}-process peak. \citet{Prantzos20} estimated that in the Solar System, these elements predominantly originate from the \textit{s} process, contributing approximately 91\%, 78\%, and 82\%, respectively, to their total abundances. 

The abundance ratios of these elements [El/Fe] exhibit an increasing trend as the metallicity [Fe/H] and [Y/Mg] decrease (see panels \textit{a}, \textit{b,} and \textit{c} of Fig.~\ref{fig:XFevsFeH_YMg}, respectively). This trend is more noticeable for zirconium. At lower metallicities, stars with planets show a decline followed by a subsequent flattening in the abundances of Sr and Y, which can be attributed to the balance between \textit{s}-process nucleosynthesis and iron production from type Ia supernovae (SNIa). In contrast, zirconium shows a noticeable increase in abundance with decreasing metallicity. Similar abundance patterns have been observed in previous studies, such as those by \citet{Mishenina16} and \citet{Delgado18}. Overall, our results for these light \textit{s}-process-dominated elements in stars with detected planets show patterns consistent with those observed in comparison stars. We also observe that younger, metal-rich dwarf stars, characterised by higher [Y/Mg] ratios, tend to show higher Sr, Y, and Zr abundances on average. 

\subsubsection{\textit{Barium, lanthanum, and cerium}}
Ba, La, and Ce are predominantly synthesised via the \textit{s} process, with their main production attributed to the so-called main component of the \textit{s} process operating in AGB stars. While a minor fraction of their production may still come from the \textit{r} process, the overall abundance pattern of Ba, La, and Ce in stellar atmospheres is expected to be strongly shaped by their common \textit{s}-process origin.

In panel \textit{d} of Fig.~\ref{fig:XFevsFeH_YMg}, we see that at super-solar metallicity, the abundance of barium increases with decreasing [Fe/H]. The \textit{s}-process (89\%; \citealt{Prantzos20}) second-peak element, Ba, at sub-solar metallicity ([Fe/H]$\leq$ --0.2) flattens and starts to decline with metallicity. A similar trend for Ba is observed in our comparison sample and in \citet{Delgado18}. Our findings do not indicate any peculiarities in the [Ba/Fe] versus [Fe/H] relationship for stars with planets. However, as shown in panels  \textit{e} and \textit{f} of Fig.~\ref{fig:XFevsFeH_YMg}, the abundances of lanthanum (80~\% \textit{s} process; \citealp{Prantzos20}) and cerium (85~\% \textit{s} process; \citealp{Prantzos20}) increase as [Fe/H] decreases. On average, stars with planets are overabundant in lanthanum and cerium at a given metallicity.

Our sample includes stars from both the thin- and thick-discs, with the majority belonging to the thin-disc. We examined whether the observed underabundance of cerium and lanthanum could be attributed to thick-disc stars. However, this is not the case; the underabundance is primarily associated with thin-disc stars. The abundance trend observed in our data may be related to a similar scenario proposed for magnesium. Enhanced magnesium abundances might compensate for the lack of iron during planet formation in metal-poor PHSs, as suggested by \cite{Adibekyan12}. It remains to be explored whether there are expected astrophysical reasons behind these relationships or if they might be coincidental. To confirm the potential overabundance of Ce and La, especially in metal-poor PHSs, further data are required.

\subsubsection{\textit{Praseodymium and neodymium}}
Both praseodymium (54\%) and neodymium (62\%) are elements formed in \textit{s} processes \citep{Prantzos20}. In panels  \textit{g} and  \textit{h} of Fig.~\ref{fig:XFevsFeH_YMg}, we see that the [Pr;~Nd/Fe] abundance ratios are increasing with decreasing metallicity [Fe/H]. The lack of strong differences in these elements in stars with and without planets suggests that their role in planet formation is more complex and likely dependent on a combination of stellar age, Galactic chemical evolution, and local enrichment conditions.

\subsubsection{\textit{Europium}}
Europium is mainly formed by the \textit{r} process \citep[95~\%;][]{Prantzos20}, and it shows an increasing trend with decreasing metallicity [Fe/H] in PHSs. As the majority of europium is produced in massive stars \citep{Matteucci21}, the [Eu/Fe] ratio is expected to decrease once iron from SNIa begins to influence the chemical evolution of the Galaxy. Consequently, metal-poor PHSs exhibit a higher [Eu/Fe] ratio on average compared to metal-rich hosts. A similar trend is observed for praseodymium \citep[47~\% \textit{r} processes;][]{Prantzos20} and neodymium \citep[39~\% \textit{r} processes;][]{Prantzos20} as both elements, similarly to europium, receive significant contributions from the \textit{r} process. 

Our analysis of \textit{n}-capture element abundances in PHSs is consistent with the Galactic chemical evolution in the solar neighbourhood. To further investigate whether \textit{n}-capture element abundances differ systematically between planet-hosting stars and comparison stars, we performed a bootstrap resampling analysis, which is described in detail in Appendix~\ref{appendix:bootstrap}. This test compared mean [El/Fe] values between the two groups while matching stars in effective temperature, surface gravity, and metallicity. The results indicate that for several elements, especially [Zr/Fe], [La/Fe], and [Ce/Fe], planet-hosting stars tend to show overabundances than comparison stars. While these differences were found consistently across bootstrap iterations, they are generally modest in magnitude and may be comparable to the typical abundance uncertainties. Therefore, we interpret these trends with caution and consider them as possible, but not definitive, chemical signatures related to planet formation.

   \begin{figure}
    \centering
    \includegraphics[width=0.9\hsize]{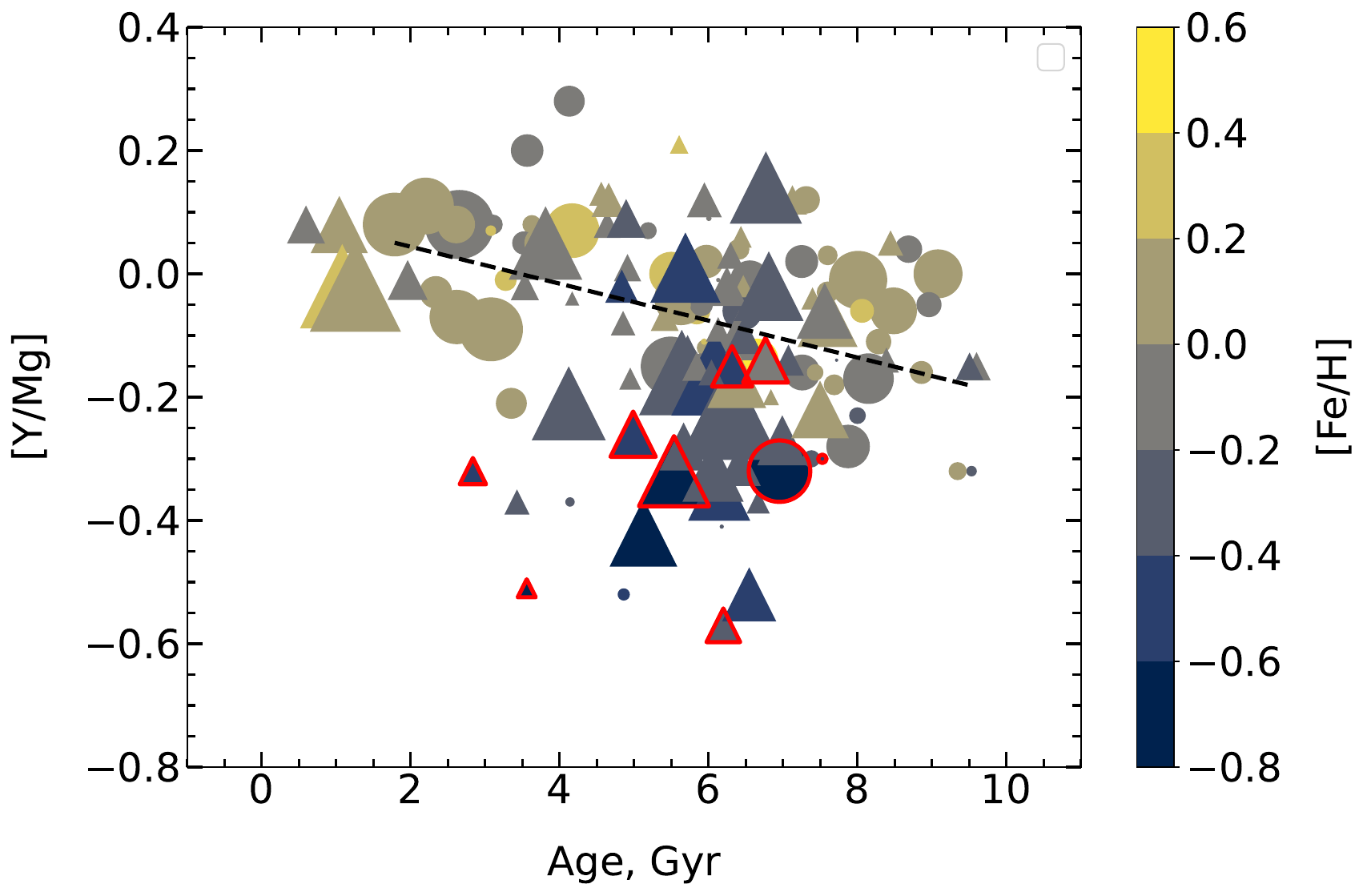}
    \caption{[Y/Mg] NLTE ratio as a function of stellar age. Stars are colour-coded by [Fe/H]. Dwarf stars are represented by circles, while giant stars are denoted by triangles. Thick-disc stars are marked with red borders. The symbol size represents the mean mass of the planets. The black line represents the least-squares linear fit for thin-disc dwarfs.}
    \label{fig:YMgvsAge_FeH}
   \end{figure}

   \begin{figure*}
    \centering
    \includegraphics[width=0.9\hsize]{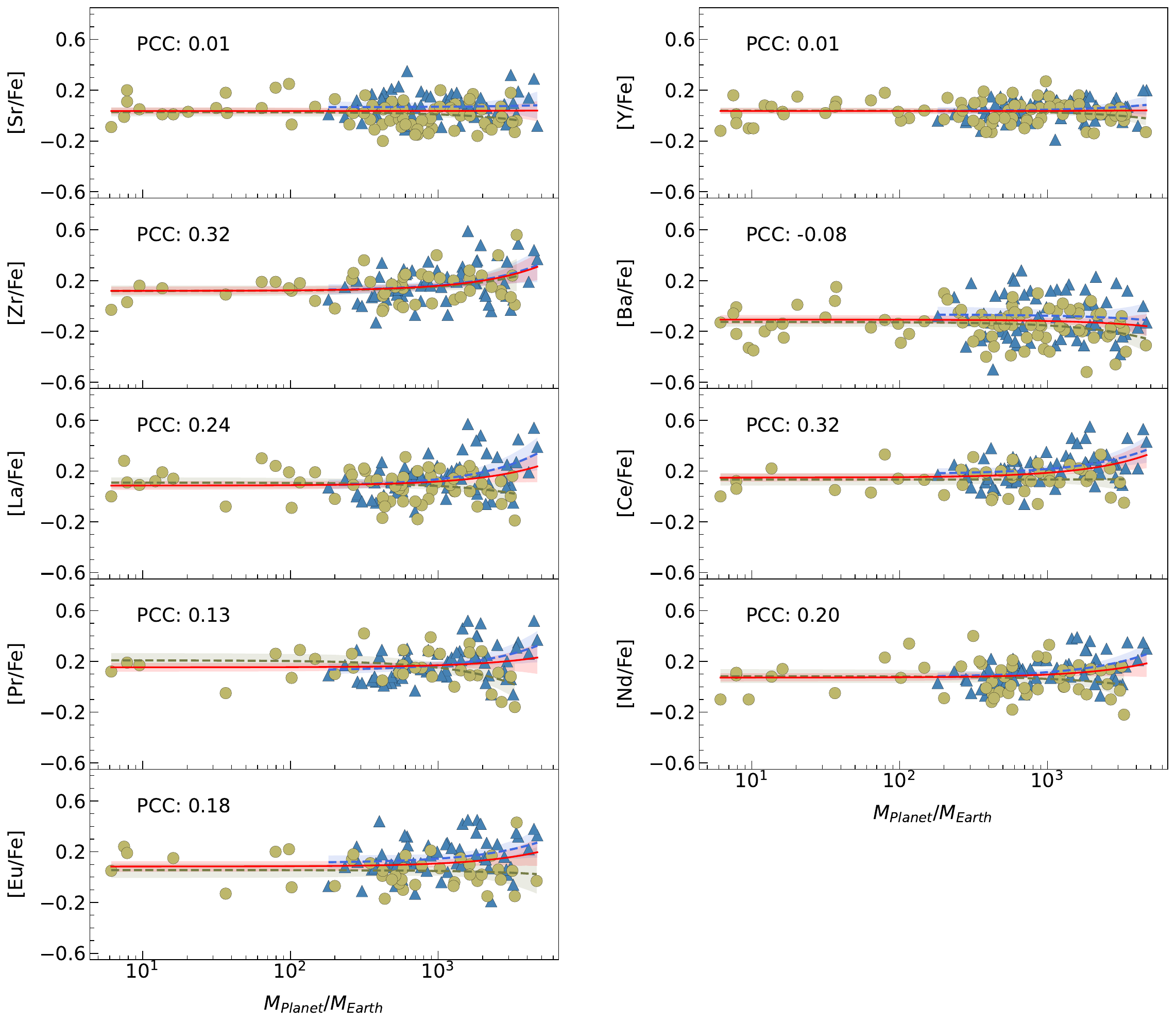}
    \caption{Abundances of neutron-capture elements [El/Fe] shown as functions of planet masses for stars with single planets, or the highest mass planet in case of multi-planetary systems. All symbols have the same meaning as in Fig.~\ref{fig:XFevsFeH_YMg}. See text for more information.}
    \label{fig:XFevsMp_HM}
   \end{figure*}

   \begin{table*}
    \caption{Pearson correlation coefficient (PCC) values representing the strength and direction of the linear relationship between stellar abundances and planet masses for both dwarf and giant stars, as well as for the entire sample.} 
    \label{table:pcc}      
    \centering
    \renewcommand{\tabcolsep}{3.5mm}
    \renewcommand{\arraystretch}{1.5}
    \begin{tabular}{l l c c c c c c}
    \hline
    \hline
    \multirow{2}{*}{Formation} & \multirow{2}{*}{[El/Fe]} & \multicolumn{3}{c}{All planets } & \multicolumn{3}{c}{Single/massive planet} \\
    \cline{3-5} \cline{6-8}
    & & All stars & Dwarfs & Giants & All stars & Dwarfs & Giants  \\
    \hline
    \multirow{4}{*}{light \textit{s}-process}       & $[\mathrm{Sr}/\mathrm{Fe}]$   &    0.05   &  $-0.08$  &    0.00   &    0.01   &  $-0.17$  &    0.03   \\
                                                    & $[\mathrm{Y}/\mathrm{Fe}]$    &    0.15   &    0.05   &    0.13   &    0.01   &  $-0.15$  &    0.14   \\
                                                    & $[\ion{Zr}/\mathrm{Fe}]$      &    0.36   &    0.31   &    0.33   &    0.32   &    0.28   &    0.32   \\
    \hline
    \multirow{3}{*}{heavy \textit{s}-process}       & $[\mathrm{Ba}/\mathrm{Fe}]$   &  $-0.03$  &  $-0.17$  &  $-0.04$  &  $-0.08$  &  $-0.21$  &  $-0.06$  \\
                                                    & $[\mathrm{La}/\mathrm{Fe}]$   &    0.27   &  $-0.10$  &    0.43   &    0.24   &  $-0.19$  &    0.41   \\
                                                    & $[\mathrm{Ce}/\mathrm{Fe}]$   &    0.40   &    0.15   &    0.39   &    0.32   &    0.01   &    0.38   \\
    \hline                                     
    \multirow{2}{*}{mixed \textit{r}+\textit{s}}    & $[\mathrm{Pr}/\mathrm{Fe}]$   &    0.15   &  $-0.26$  &    0.36   &    0.13   &  $-0.41$  &    0.34   \\
                                                    & $[\mathrm{Nd}/\mathrm{Fe}]$   &    0.19   &  $-0.09$  &    0.38   &    0.20   &  $-0.14$  &    0.38   \\
    \hline
    \multirow{1}{*}{\textit{r}-process}             & $[\mathrm{Eu}/\mathrm{Fe}]$   &    0.21   &  $-0.07$  &    0.25   &    0.18   &  $-0.06$  &    0.26   \\
    \hline
    \hline
    \end{tabular}

   \end{table*}

\subsubsection{\textit{The [Y/Mg]} clock}
Recent studies have explored the relationship between elemental abundances and stellar age, revealing that the ratio of \textit{s}-process elements (e.g. Y) to $\alpha$ elements (e.g. Mg) is linearly correlated with stellar age \citep[e.g.][]{Nissen15, Feltzing17, Slumstrup17, Tautvaisiene21, Storm23, Shejeelammal24}. Chemical abundances are intrinsically linked to a stellar age and place of origin, making them a valuable tool for studying the stars and the evolutionary history of the Milky Way. The correlation between [Y/Mg] and age arises from the different nucleosynthesis timescales for these elements. \textit{s}-process elements are mainly produced during the AGB phase of low- and intermediate-mass stars, which occurs on a longer timescale, whereas $\alpha$-elements are mainly synthesised during type II supernovae (SNII), which occur on much shorter timescales. 

Figure~\ref{fig:YMgvsAge_FeH} shows the relationship between the NLTE [Y/Mg] ratio and stellar age for our sample. Dwarf stars are marked as circles, while giant stars are represented by triangles. Stars are colour-coded according to their [Fe/H] values, with thick-disc members outlined in red. The size of each symbol reflects planetary mass specifically, the average mass in multi-planetary systems, and the individual mass in single-planet systems. Stellar ages and magnesium abundances are derived from our previous work \citepalias{Sharma24}. 

The [Y/Mg]-age correlation is present for thin-disc dwarf stars hosting planets with  a [Y/Mg]–age slope of $-$0.030~$\pm$~0.008. This result is consistent with the slope derived in thin-disc solar-neighbourhood stars, as reported by \cite{Tautvaisiene21}. Additionally, metal-poor thick-disc stars exhibit enhanced magnesium abundances and lower [Y/Mg] ratios. Mg might compensate for the lack of iron in forming planets in metal-poor thick-disc PHSs \citep[see e.g.][]{Adibekyan12, Adibekyan12b, Bashi19}.

\subsection{Stellar chemical composition and planet mass relation}
\label{subsec:planetmass}
In this section, we investigate the correlation between \textit{n}-capture element abundances in PHSs and the masses of their orbiting planets to explore potential links between stellar chemistry and planet formation. We adopted the methodology from our previous study to classify our sample of host stars into dwarfs and giants. The planetary masses used in this study were primarily sourced from literature, mainly the NASA Exoplanet Archive. The parameters for 149 star-planet systems were previously published in \citet{Tautvaisiene22} and \citetalias{Sharma24}, while those for 11 newly identified systems are summarised in Table~\ref{table:exoplanets}. To minimise potential biases arising from stellar evolution, we identified and excluded 15 planets with masses exceeding approximately 13 M$_{Jupiter}$ from our analysis. This mass threshold corresponds to the minimum mass required for the thermonuclear fusion of deuterium, indicating that such objects may be classified as brown dwarfs rather than planets.

Figure~\ref{fig:XFevsMp_HM} shows the distribution of \textit{n}-capture element abundances relative to iron as a function of planet masses for the investigated dwarf and giant stars. For multi-planetary systems, only the most massive planet was considered. However, we also analysed the elemental abundance trends with planet mass by incorporating the masses of all planets within these systems. Our analysis revealed no significant differences when including the full set of planetary masses within these systems (see Fig.~\ref{fig:XFevsMp}).

As in \citetalias{Sharma24}, we performed a linear-regression analysis to explore a potential correlation between element abundances and planet masses. We computed the linear fits separately for dwarf and giant stars, as well as the entire sample. We also assessed the strength and direction of the linear relationships by calculating the Pearson correlation coefficient (PCC) values for each element. These values are presented in Table~\ref{table:pcc} and detailed below.
\paragraph{The first \textit{s}-process peak: light \textit{s}-process elements Sr, Y, Zr.} Our results show a very weak correlation for strontium with planetary mass across all cases, with PCC values ranging from –0.17 for dwarfs with single or the most massive planets to 0.01 for the full sample. This suggests that Sr abundance is largely independent of planet mass, although a weak depletion trend may still be marginally present in dwarf stars with massive planets. \citet{Delgado18} showed similar results for Sr, underscoring a mild correlation of Sr abundance with planetary mass. For yttrium, we find a weak correlation of [Y/Fe] with planet mass for all cases, which is consistent with the lack of a strong trend reported by \cite{Swastik22} for main-sequence stars. Zirconium shows a consistently positive correlation towards high-mass planets across all cases, suggesting that Zr is more enriched in host stars with more massive planets. This is in contrast to the negative trend reported by \citet{Swastik22}, and the discrepancy may arise from differences in the adopted Zr\,{\sc ii} spectral lines or subtle influences from Galactic chemical evolution. Notably, the higher Zr abundances observed in giants hosting massive planets can be predominantly associated with marginally older stars (see Fig.~\ref{fig:XFevsMp_YMg}).

\paragraph{The second \textit{s}-process peak: heavy \textit{s}-process elements Ba, La, and Ce.} Our analysis of barium abundance as a function of planet mass for target stars reveals a weak negative correlation across all cases, with PCC values ranging from –0.03 to –0.08, most notably in dwarf stars with single or massive planets. This suggests a mild depletion of Ba in stars hosting more massive planets, although the trend remains weak. Lanthanum and cerium show moderate-to-strong correlations with planetary mass, particularly in giant stars, where the PCCs reach 0.43 and 0.39, respectively. These consistent positive trends may indicate a potential link between the enrichment of these heavy \textit{s}-process elements and the formation or presence of massive planets.

\paragraph{Mixed elements Pr and Nd.} Praseodymium and neodymium follow the same strong correlation with planet mass as observed for La and Ce in giant stars, with PCC values reaching 0.36 and 0.38, respectively. In dwarf stars, Pr shows a strong negative correlation, especially in stars with single or massive planets. 

\paragraph{The \textit{r}-process-dominated element Eu.} Europium exhibits a weak positive correlation with planet mass in the overall and giant star samples, which broadly consistent with the trend observed for La and Ce. However, the strength of the correlation is notably lower, and in dwarf stars the correlation is slightly negative, suggesting a weaker relationship between this pure \textit{r}-process element and planetary mass.

Our findings suggest a potential link between the abundances of \textit{n}-capture elements in giant stars and the planetary mass. In particular, several elements, including zirconium, lanthanum, cerium, praseodymium, neodymium, and europium, exhibit moderate-to-strong positive correlations with higher mass planets in giant stars. However, exceptions include strontium, yttrium, and barium, which consistently show negligible or weak correlations across all cases. The interpretation of these trends remains complicated by the effects of Galactic chemical evolution. This can obscure or even erase chemical signatures associated with planet formation, making it difficult to distinguish between planetary influences and broader evolutionary trends.

\subsection{Elemental abundances versus condensation temperatures}
\label{subsec:condensationTemperatura}

   \begin{figure}
    \centering
    \includegraphics[width=0.8\hsize]{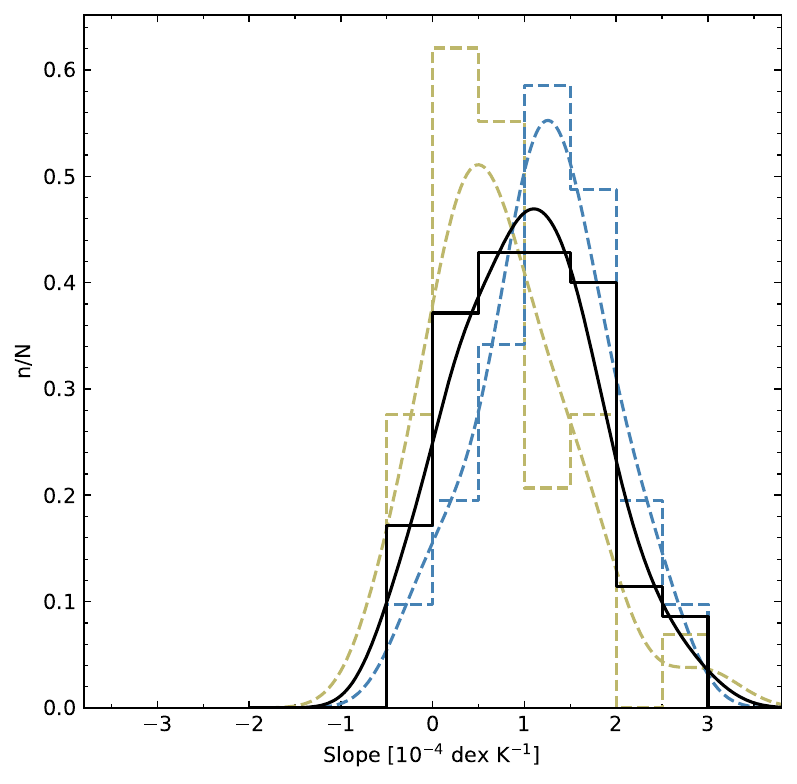}
    \caption{Distributions of $\Delta$[El/H]–$T_{\text{cond}}$ slopes for our stars with planets after abundance corrections with comparison star(s). The solid black line represents all planet hosts, while the yellow and blue dashed lines represent dwarf and giant stars.}
    \label{fig:Tcon}
   \end{figure}

   \begin{figure*}
    \centering
    \includegraphics[width=\hsize]{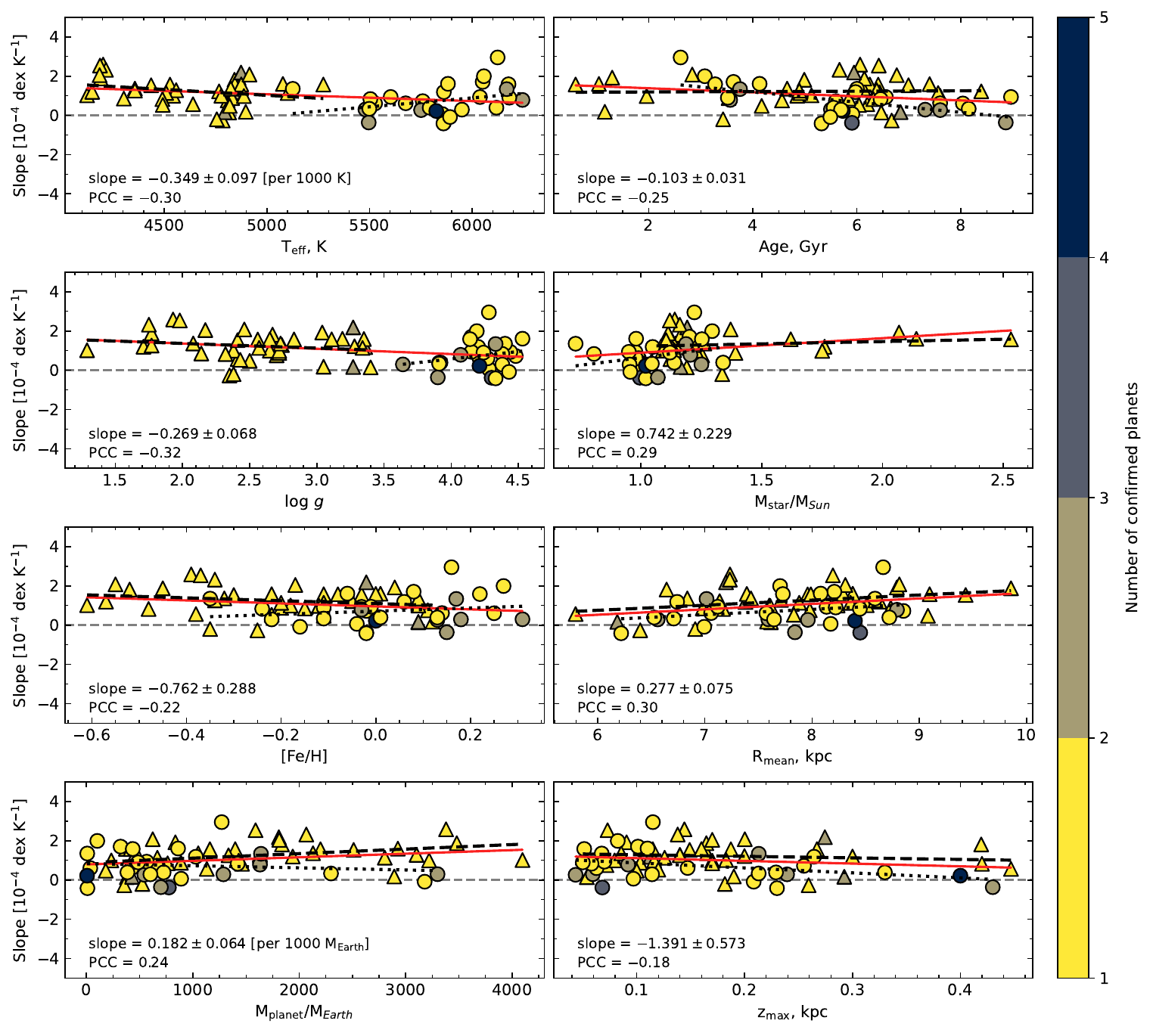}
    \caption{$\Delta$[El/H]–$T_{\text{cond}}$ slopes as a function of stellar parameters and planetary mass. The stars are colour-coded according to their number of confirmed planets. The symbols are the same as in Fig.~\ref{fig:XFevsFeH_YMg}. The red line represents weighted linear regression fits to the data, while the black dotted and dashed lines correspond to dwarf and giant star hosts, respectively. The slopes and Pearson correlation coefficients are shown in the corners of the panels. See the text for more information.}
    \label{fig:Slopes_vs_params}
   \end{figure*}
  
The relationship between elemental abundance differences $\Delta[\text{El}/\text{Fe}]$ in PHSs and their comparison counterparts as a function of condensation temperature ($T_{\text{cond}}$) remains an active area of debate. Several studies have reported variations in the abundance patterns of volatile and refractory elements in planet hosts. 

In their pioneering study, \cite{Melendez09} found that the Sun exhibits a depletion in refractory elements (those with high $T_{\text{cond}}$) compared to volatile elements (those with low $T_{\text{cond}}$) when compared with solar twins. This pattern suggests that refractory elements were preferentially incorporated into planetesimals and planets during the early stages of the Solar System's formation, leaving the Sun relatively deficient in these elements.

Extending this analysis to other planetary systems, \cite{Gonzalez13} conducted a comprehensive analysis of volatile-to-refractory abundance ratios in a sample of 61 F- and G-type stars, including 29 planet hosts. Their abundance trends with condensation temperature displayed nearly null or slightly negative slopes. They revisited the sample of solar analogues, focusing on eight stars with super-Earth-like planets and found that only four of these stars showed clear increasing abundance trends with $T_{\text{cond}}$. This suggests that the volatile-to-refractory abundance ratio may not be directly related to the presence of rocky planets.

\citet{Mack14} analysed the HD 20782/81 wide binary system where both stars host giant planets and showed that both stars exhibit positive correlations for refractory elements ($T_{\text{cond}}$>900~K) between elemental abundances (relative to solar) and $T_{\text{cond}}$, suggesting the ingestion of rocky material during planet formation. In another work, \cite{Liu20} studied a sample of 16 PHSs and 68 comparison stars, finding diverse abundance trends as a function of $T_{\text{cond}}$. While some stars displayed positive correlations, others showed negative or no significant trends. 

Recently, \cite{Yun24} refined these interpretations by analysing 227 stars with planets, focusing on abundance trends of refractory elements. The study examined six refractory elements (Mg, Si, Ca, Al, Mn, and Ni), demonstrating that while an [El/Fe] versus $T_{\text{cond}}$ trend exists, it depends on the specific planetary architecture of the system. After accounting for Galactic chemical evolution, they found that stars hosting giant planets are more depleted in refractory elements compared to those hosting only rocky planets. This suggests different material distributions in protoplanetary discs with massive planets compared to those with only rocky planets. Additionally, among stars hosting only rocky planets, the study identified correlations between refractory depletion trends and planetary parameters, such as planet radius and multiplicity, indicating that the composition of a stellar system may be shaped by the properties of its planets. These findings further support the idea that planet formation processes leave detectable imprints on stellar compositions and highlight the importance of considering planetary system architecture when interpreting abundance trends.

To further explore any potential stellar chemical signature of planet formation, we calculated the abundance difference versus condensation temperature for our stars with planets, subtracting the mean abundance values from comparison stars with similar parameters (the difference allowed in the main atmospheric parameter values: $T_{\rm eff}$~$\pm$75~K;~${\rm log}~g$~$\pm$0.25; [Fe/H]~$\pm$0.11). We also took into account thin- or thick-disc attribution. Up to 23 comparison stars were found for each PHS, with six on average that have analysed elemental abundances. The large comparison sample helps mitigate potential Galactic and stellar chemical-evolution effects and biases, utilising homogeneous results from our previous studies. Additionally, with \textit{n}-capture element abundances, we included the abundances of C, N, O, Mg, and Si elements from \citetalias{Sharma24} and our earlier works \citep{Stonkute20, Tautvaisiene22}. Condensation temperatures were taken from the 50\% $T_{\text{cond}}$ values derived by \citet{Lodders03} for a Solar System gas with [Fe/H]= 0. All our \textit{n}-capture elements, along with Mg and Si, are refractory elements with high condensation temperatures ($T_{\text{cond}}$ > 1300 K). In contrast, C, N, and O have much lower condensation temperatures ($T_{\text{cond}}$ < 180 K) and are volatiles.

In Fig.~\ref{fig:Tcon}, we present the calculated distributions of the C, N, O, Mg, Si, Sr, Y, Zr, Ba, La, Ce, Nd, Pr, and Eu abundance differences ($\Delta$[El/H]) in the star with planets minus comparison stars, versus $T_{\text{cond}}$ slopes. For each PHS, we identify a set of comparison stars and examine the abundances corrected for the Galactic and stellar chemical evolution effects, as a function of the $T_{\text{cond}}$ of the individual elements. This figure uses the $T_{\text{cond}}$ slopes derived  considering refractory and volatile elements. If a star with a planet(s) lacks the determination of volatile elements, we exclude it from the general distribution. The yellow and blue dashed lines indicate dwarf and giant stars, respectively, while the solid black line represents the whole sample. The overall $\Delta$[El/H]–$T_{\text{cond}}$ slope distribution shows a positive skewness ($\mu_{1/2} = 1.01$~$\pm$ 0.75~x~10$^{-4}$ dex K$^{-1}$) for all planet hosts (black line), indicating that most of the stars with planets are refractory rich compared to the comparison sample. When considering only the dwarf star sample, the results are closer to zero ($\mu_{1/2} = 0.67$~$\pm$ 0.75~x~10$^{-4}$ dex K$^{-1}$), suggesting less significant abundance differences versus $T_{\text{cond}}$ correlation. Our analysis of a large sample of stars with planets reveals a diversity of abundance–$T_{\text{cond}}$ slopes, which aligns with previous studies \citep[e.g.][]{Mishenina16, Liu20, Tautvaisiene22}.

The variations observed in the abundance-$T_{\text{cond}}$ slopes might be influenced by different stellar and planetary parameters. Figure~\ref{fig:Slopes_vs_params} shows the dependencies of the $\Delta$[El/H]–$T_{\text{cond}}$ slopes on these parameters. The red line represents weighted linear regression fits to the entire dataset, while the black dotted and dashed lines correspond to dwarf and giant-star hosts, respectively. The inverse of the variance ($\sigma$${^2}$) of the slopes was used as the weight in the linear regression. As illustrated, there are some correlations between the abundance slopes and stellar parameters or planetary mass, but they are not very significant. Nevertheless, we note that older dwarf stars hosting two or more planetary companions tend to exhibit smaller or even negative  $\Delta$[El/H]–$T_{\text{cond}}$ slopes on average compared to younger dwarfs (Age < 5~Gyr) which have positive slopes in our sample. Also, metal-rich dwarf stars ([Fe/H]>0) tend to show higher $\Delta$[El/H]–$T_{\text{cond}}$ slope values relative to their more metal-poor counterparts. Our results also show that multi-planetary systems are more common around metal-rich stars.

The variety of the observed $\Delta$[El/H]–$T_{\text{cond}}$ slopes found for planet-hosting stars might be associated with various processes, such as a variety of evolutionary paths of circumstellar discs, accretion or formation of terrestrial planets, and giant-planet rocky cores. However, as noted in previous studies, we cannot dismiss the possibility that there is no physical connection between the presence of planets and the $T_{\text{cond}}$ trend. An alternative explanation proposed by \cite{Soliman25} suggests that fluctuations in refractory surface element abundances can result from variations in dust-to-gas ratios near stars during their formation. This mechanism indicates that these abundance variations would persist throughout a star's lifetime, including its earlier stages with thicker convective zones, potentially leading to both reductions and enhancements in refractory surface abundances.

\section{Summary and conclusions}
\label{sec:conclusions}
This paper investigates the chemical composition of planet-hosting stars (referred to in this paper as stars with planets), focusing on \textit{n}-capture elements to understand their role in planet formation and stellar evolution. Using high-resolution spectroscopy, we analysed the elemental abundances of a selected sample of 160 stars with planets, including 86 dwarf stars and 74 giant stars, to examine correlations between heavy elements and planetary presence. This study builds upon our previous research on C, N, and $\alpha$-elements in exoplanet host stars \citepalias{Sharma24}, extending the analysis to elements formed through slow- (\textit{s}-process), rapid- (\textit{r}-process), and mixed-process \textit{n}-capture mechanisms. By studying stars with planets, we aim to explore possible signatures of planet formation imprinted in stellar chemical abundances.

Observational data were obtained using the high-resolution Vilnius University Echelle Spectrograph (VUES) at the Mol\.{e}tai Astronomical Observatory. The main atmospheric parameters were derived in \citetalias{Sharma24} using the classical equivalent-width approach. Detailed abundance measurements of \textit{n}-capture elements were performed using the spectral synthesis method. The abundance patterns of these elements relative to iron were then analysed across a wide range of metallicities.

The detailed abundances of exoplanet host stars are mainly consistent with the Galactic chemical evolution. The results indicate that light \textit{s}-process elements, Sr, Y, and Zr exhibit an increasing abundance ratio of [El/Fe] as the metallicity and [Y/Mg] decrease at solar [Fe/H] values, with the trend being more pronounced for zirconium. However, we observed a decline in Sr and Y at lower metallicities, while Zr follows an increasing pattern. Additionally, younger, metal-rich dwarf stars with higher [Y/Mg] ratios exhibited enhanced Sr, Y, and Zr abundance. 

For the second-peak \textit{s}-process elements Ba, La, and Ce, we found that while Ba remained relatively flat at sub-solar metallicities, La and Ce exhibited increasing trends with decreasing metallicity. On average, stars with planets showed an overabundance of La and Ce at a given metallicity. 

The heavier \textit{n}-capture elements Pr and Nd, which receive contributions from both \textit{s} and \textit{r} processes, displayed moderately increasing trends with decreasing metallicity. Eu, predominantly produced via the \textit{r} process, followed a similar trend, with metal-poor stars exhibiting higher [Eu/Fe] ratios, which is consistent with its origin in massive stars and subsequent enrichment from supernovae.

We also analysed the relationship between \textit{n}-capture element abundances in stars with planets and the masses of their planetary companions. Our results lead to the conclusions listed below.

\begin{itemize}
 \item \textit{s} process: Sr, Y, and Ba exhibit no significant correlation with planet mass. Zr displays a positive correlation in all cases, which may be linked to the age distribution of the host stars or subtle Galactic chemical-evolution effects. La and Ce display moderate positive correlations, particularly in giant stars, suggesting a potential link between these elements and planetary formation processes.

 \item \textit{r+s} processes: the mixed elements Pr and Nd follow similar trends to La and Ce, alongside the well-established role of iron, which reinforces the idea that certain \textit{n}-capture elements may be linked to the presence and characteristics of planetary companions.

 \item \textit{r} process: Eu shows a weak positive correlation with planetary mass. This trend aligns with those observed for other \textit{n}-capture elements, though the statistical significance remains limited.
\end{itemize}

The Galactic chemical evolution, responsible for the chemical diversity among stars formed at different times of the Milky Way, can obscure or even erase potential planetary signatures, further complicating efforts to determine whether observed differences are due to planets or to evolutionary trends.

Furthermore, we explored how the abundance differences between our stars with planets and carefully selected comparison stars vary as a function of elemental condensation temperature. Alongside \textit{n}-capture elements, abundances of C, N, O, Mg, and Si were included for a comprehensive analysis across both volatile and refractory elements. The comparison stars from our previous work were selected based on similarity of atmospheric parameters, kinematics, and ages to minimise the effects of Galactic chemical evolution and observational biases. We found a positive skew in the $\Delta$[El/H]–$T_{\text{cond}}$ slope distribution for the overall sample of planet-hosting stars. This suggests an enrichment of refractory elements (those with high $T_{\text{cond}}$) relative to volatiles in PHSs compared to their stellar analogues. However, when looking at the dwarf-star sub-sample, the slope distribution shifts closer to zero, suggesting less significant trends in abundance differences versus $T_{\text{cond}}$. We further investigated the correlations between $\Delta$[El/H]–$T_{\text{cond}}$ slopes and various stellar and planetary parameters. While no strong correlations were found, some trends are noteworthy. On average, older dwarf stars hosting multiple planets tend to display lower (or even negative) $\Delta$[El/H]–$T_{\text{cond}}$ slopes, and metal-rich dwarf stars show higher (more positive) abundance–$T_{\text{cond}}$ slopes compared to their more metal-poor counterparts. Our results also show that multi-planetary systems are more common around metal-rich stars.

Finally, the results presented in this study serve as an important benchmark for future investigations. By offering a detailed chemical composition of \textit{n}-capture elements in planet-host stars, this work contributes to a growing effort to link stellar chemical compositions with the architecture and formation history of planetary systems. These results hold significance not only for Galactic evolution studies but also as useful priors for further exoplanet modelling. Ultimately, understanding these chemical fingerprints is key to unravelling the broader narrative of planet formation within the Galaxy. 

\vspace{2mm}
\textit{Data availability}: Table~\ref{table:Results} is only available in electronic form at the CDS via anonymous ftp to cdsarc.u-strasbg.fr () or via \url{http://cdsweb.u-strasbg.fr/cgi-bin/qcat?J/A+A/}.

\begin{acknowledgements}
We thank the anonymous referee for their helpful and constructive suggestions. E.S., A.D., R.M., {\v S}.M. and G.T. acknowledge funding from the Research Council of Lithuania (LMTLT, grant No. P-MIP-23-24). This research has made use of the NASA Exoplanet Archive, which is operated by the California Institute of Technology, under contract with the National Aeronautics and Space Administration under the Exoplanet Exploration Programme. We also acknowledge the use of the SIMBAD database, operated at CDS, Strasbourg, France. U.J. acknowledge funding from the Research Council of Lithuania (LMTLT, grant No. S-LL-24-1). Additionally, we appreciate the Vilnius University's Mol\.{e}tai Astronomical Observatory for granting us observation time for this project. The observing time was partially funded by the Europlanet Telescope Network programme of the Europlanet 2024 Research Infrastructure project. Europlanet 2024 RI has received funding from the European Union's Horizon 2020 research and innovation programme under grant agreement No 871149.
\end{acknowledgements}

\bibliographystyle{aa}
\bibliography{aa55466-25}

\appendix
\onecolumn
\section{Statistical tests} \label{appendix:bootstrap}
We carried out a detailed statistical analysis to compare the \textit{n}-capture abundances ([El/Fe]) of our sample of planet-hosting stars with those of a control sample of stars from our previous work \citep{Tautvaisiene21}. To ensure meaningful comparisons, the analysis was performed separately for dwarf and giant stars, taking into account differences in stellar structure and evolutionary stages. For each planet-hosting star, on average, five comparison stars were selected from the control sample with similar stellar parameters (within $\pm100$~K in effective temperature, $\pm0.2$~dex in surface gravity, and $\pm0.1$~dex in metallicity). This careful matching helps isolate the effect of planetary presence from the underlying stellar characteristics.

To evaluate abundance differences between the two groups, we employed a bootstrap resampling technique with 10\,000 iterations, where in each iteration, one comparison star was randomly selected for each PHS from its corresponding set, and the mean abundance difference $\langle{\rm [El/Fe]}\rangle$ between the PHS and their selected comparison stars was calculated. This iterative approach allowed us to assess the statistical significance and robustness of any observed abundance differences. By repeating this process thousands of times, we can account for the variability introduced by the random sampling and ensure that any trends we observe are not driven by outliers or particular subsets of stars. Ultimately, this method enables us to assess whether the presence of planets is associated with statistically significant differences in \textit{n}-capture element abundances, beyond what could be expected from random variations or parameter mismatches alone.

   \begin{table}[h!]
    \caption{Bootstrap resampling test results for \textit{n}-capture elemental ratio distributions for stars with and without planetary companions.} 
    \label{table:bootstrap}      
    \centering
    \renewcommand{\arraystretch}{1.35}
    \begin{tabular}{l c c c c}
    \hline
    \hline
    & \multicolumn{2}{c}{Dwarfs} & \multicolumn{2}{c}{Giants} \\
    \cline{2-3} \cline{4-5}
    & Difference ($\Delta$) & $p$-value & Difference ($\Delta$) & $p$-value \\ 
    \hline  
    $[\mathrm{Sr}/\mathrm{Fe}]$ & $-0.02$ &       0.012   & $-0.08$ & $\leq{0.001}$ \\   
    $[\mathrm{Y}/\mathrm{Fe}]$  &   0.04  & $\leq{0.001}$ &   0.09  & $\leq{0.001}$ \\
    $[\mathrm{Zr}/\mathrm{Fe}]$ &   0.14  & $\leq{0.001}$ &   0.14  & $\leq{0.001}$ \\
    $[\mathrm{Ba}/\mathrm{Fe}]$ & $-0.02$ &       0.008   &   0.06  & $\leq{0.001}$ \\
    $[\mathrm{La}/\mathrm{Fe}]$ &   0.12  & $\leq{0.001}$ &   0.10  & $\leq{0.001}$ \\
    $[\mathrm{Ce}/\mathrm{Fe}]$ &   0.13  & $\leq{0.001}$ &   0.24  & $\leq{0.001}$ \\
    $[\mathrm{Pr}/\mathrm{Fe}]$ &   0.03  &       0.052   & $-0.02$ &       0.015   \\
    $[\mathrm{Nd}/\mathrm{Fe}]$ &   0.07  & $\leq{0.001}$ &   0.01  &       0.171   \\
    $[\mathrm{Eu}/\mathrm{Fe}]$ &   0.06  & $\leq{0.001}$ &   0.08  & $\leq{0.001}$ \\
    \hline
    \end{tabular}
    \tablefoot{The test is conducted across 10\,000 iterations.} 
   \end{table}

Table~\ref{table:bootstrap} shows the mean abundance differences ($\Delta = \langle[\mathrm{X}/\mathrm{Fe}]\rangle_{\mathrm{host}} - \langle[\mathrm{X}/\mathrm{Fe}]\rangle_{\mathrm{comp}}$) between planet-hosting stars and comparison stars separately for dwarf and giant stars. Two-tailed $p$-values are also reported in the table. Results from this test showed several systematically different [El/Fe] abundance ratios compared to the control sample. For planet-hosting dwarfs, Y, Zr, La, Ce, Nd, and Eu exhibited positive mean differences ($\Delta$) with bootstrap $p$-values $\leq{0.001}$. The largest $\Delta$ values were found for Zr ($\Delta = +0.14$~dex), Ce ($\Delta = +0.13$~dex), and La ($\Delta = +0.12$~dex). Eu, an \textit{r}-process element, also showed a moderate increase ($\Delta = +0.06$~dex) hinting at a potential enrichment of \text{n}-capture elements in planet-hosting environments. Interestingly, Ba was significantly lower in planet-host dwarfs $(\Delta = -0.02\ \mathrm{dex},\ p = 0.008)$ compared to other elements, consistent with previous studies (e.g., \citep{Delgado18}), though the absolute offset is small. Sr also showed a slight depletion $(\Delta = -0.02\ \mathrm{dex},\ p = 0.012)$, and Pr showed a modest increase $(\Delta = +0.03\ \mathrm{dex},\ p = 0.052)$.

A similar pattern was observed for giant stars, where Y, Zr, La, Ce, and Eu showed higher [El/Fe] ratios in planet-hosting giants compared to their control group. Notably, cerium showed the most substantial difference among all elements ($\Delta = +0.24$~dex). Similar to dwarfs, Sr was found to be lower in planet-host giants $(\Delta = -0.08\ \mathrm{dex})$. However, Ba, unlike in dwarfs, appeared slightly higher in giants with planets $(\Delta = +0.06\ \mathrm{dex})$. This may reflect different evolutionary or internal mixing processes, or differences in the chemical environment during planet formation. Nd did not show clear difference between hosts and comparison stars ($\Delta = 0.01$~dex, $p = 0.17$) while Pr showed a small depletion $(\Delta = -0.02\ \mathrm{dex},\ p = 0.015)$.

   \begin{figure*}[h!]
    \centering
    \includegraphics[width=1\hsize]{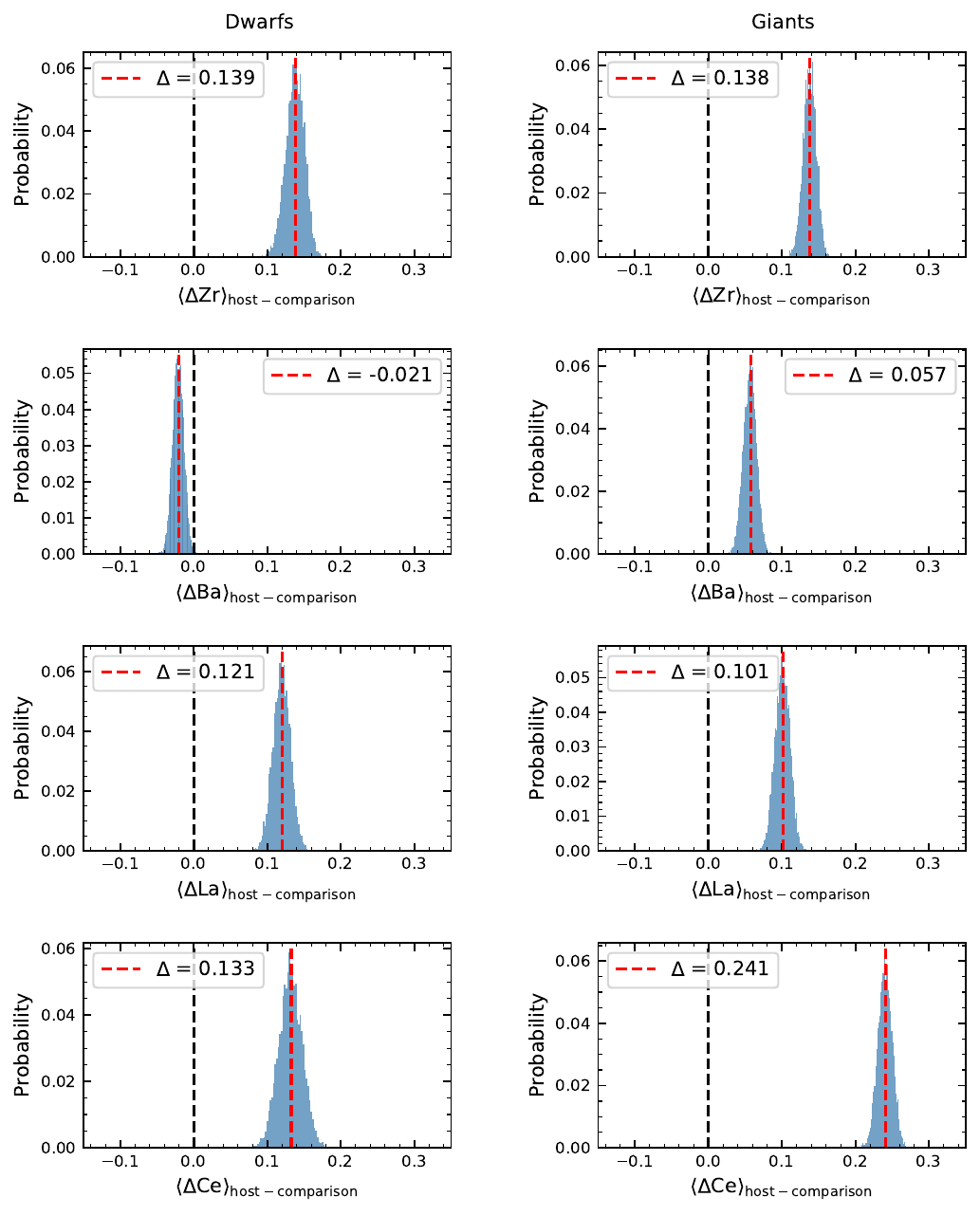}
    \caption{Bootstrap distributions of mean abundance differences ($\Delta = \mathrm{[El/Fe]_{host} - [El/Fe]_{comparison}}$) for selected \textit{n}-capture elements (Zr, Ba, La, and Ce) in dwarfs and giants. Each histogram represents the result of 10\,000 resampled comparisons with a control sample matched in $T_{\mathrm{eff}}$, $\log g$, and [Fe/H]. The vertical dashed lines mark $\Delta = 0$ (black) and the calculated mean $\Delta$ (red).}
    \label{fig:bootstrap}
   \end{figure*}

To illustrate the results of our bootstrap approach, we present example histograms of the bootstrap $\Delta$ distributions for selected elements in Figure~\ref{fig:bootstrap}. These include Zr, Ba, La and Ce, separately for dwarf and giant stars. For each element, the distribution shows the shift between the planet-host sample and the matched comparison sample. In most cases, the mean $\Delta$ is offset from zero, with Zr, La and Ce showing clear positive shifts, while Ba demonstrates a negative shift in dwarfs and a positive shift in giants.

\newpage

\section{Supplementary figures and data tables}
\begin{figure*}[h!]
    \centering
    \includegraphics[width=0.9\hsize]{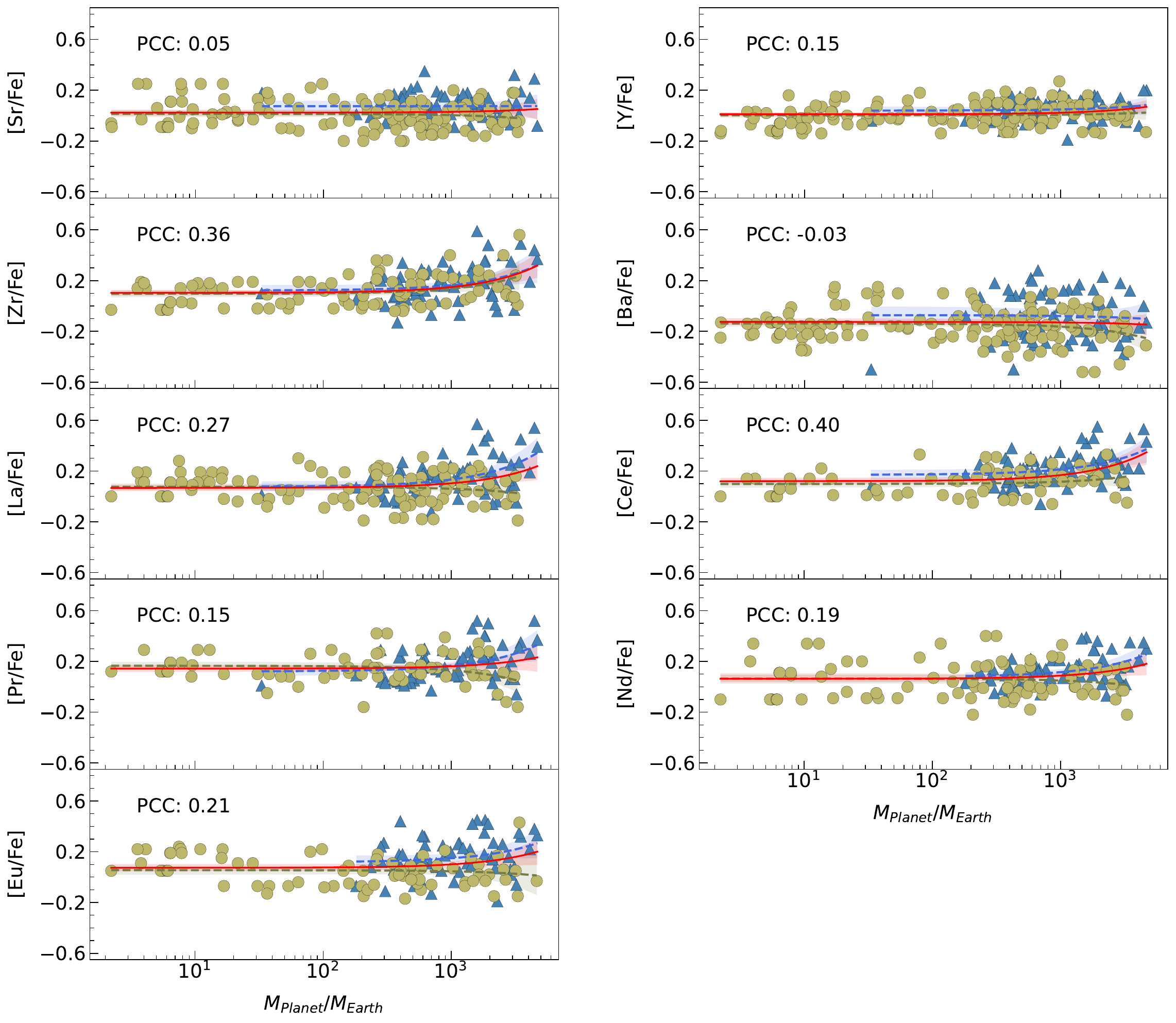}
    \caption{Abundances of neutron-capture elements [El/Fe] shown as functions of planet masses for all investigated planetary systems including all planets. All symbols have the same meaning as in Fig.~\ref{fig:XFevsFeH_YMg}. Refer to the text for more information.}
    \label{fig:XFevsMp}
   \end{figure*}

\begin{figure*}[h!]
    \centering
    \includegraphics[width=1\hsize]{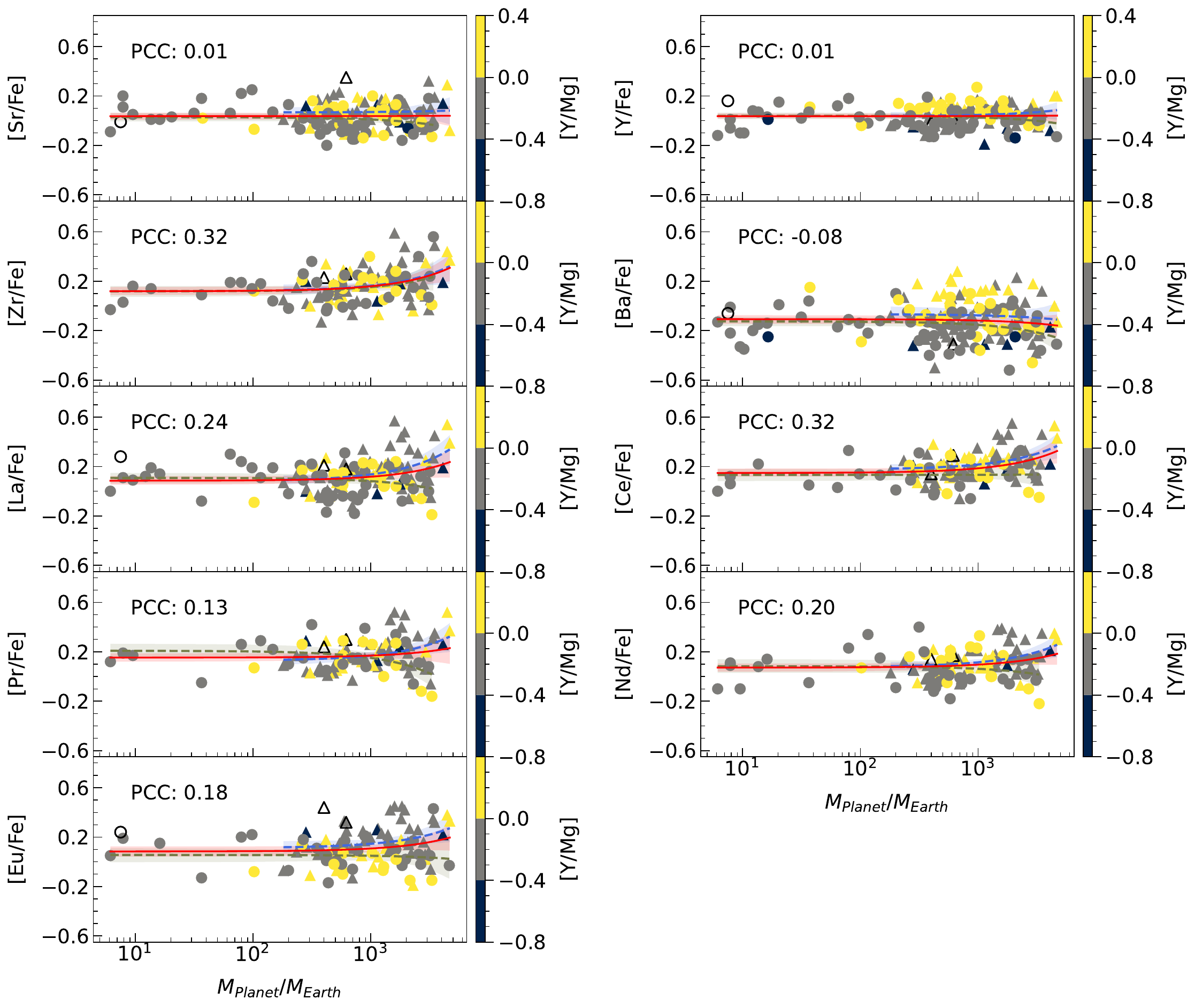}
    \caption{Abundances of neutron-capture elements [El/Fe] shown as functions of planet masses for stars with single planets or the highest-mass planet in multiplanetary systems. The data is colour-coded by [Y/Mg]. All symbols have the same meaning as in Fig.~\ref{fig:XFevsFeH_YMg}.}
    \label{fig:XFevsMp_YMg}
   \end{figure*}

Figure~\ref{fig:XFevsMp} shows the abundance ratios of \textit{n}-capture elements ([El/Fe]) plotted as a function of the planetary masses for all the confirmed planets orbiting the stars in our sample. In Fig.~\ref{fig:XFevsMp_YMg}, we explore possible correlations between the planetary mass and \textit{n}-capture element abundances for stars with single planets or the highest-mass planet in the case of multiplanetary systems. In this figure, we have colour-coded the data points according to the [Y/Mg] abundance ratio, which has been widely used as a chemical clock. Empty symbols represent stars for which magnesium abundances were not available in \citetalias{Sharma24}.

Table~\ref{table:Results} provides a comprehensive summary of the results obtained from our chemical abundance analysis. The table lists the derived chemical abundance ratios of \textit{n}-capture elements [El/H] for each star in our sample, along with their associated measurement uncertainties. For readers and researchers interested in exploring the complete dataset, the complete version of the table is made available in a machine-readable form at CDS.

In Table~\ref{table:exoplanets}, we present the key planetary parameters for the 11 newly added stars investigated in this study. The table lists important characteristics such as the planetary mass (expressed relative to the Earth's mass), the orbital period, and the orbital semi-major axis. For completeness, it should be noted that the planetary parameters for the remaining stars analysed in this work were already published and discussed in detail in \citetalias{Sharma24}.

\onecolumn 

 \renewcommand{\arraystretch}{1.21}
 \begin{longtable}{llll}
 \caption{Contents of the machine-readable table available at the CDS.}
  \label{table:Results}\\
  \hline
  \hline
  Col & Label & Units & Explanations\\
  \hline
  \noalign{\smallskip}
  1     & Host TYC ID                         & ---          & Tycho-2 catalogue identification\\
  2     & Teff                                & K            & Effective temperature\\
  3     & e\_Teff                             & K            & Uncertainty in effective temperature\\
  4     & Logg                                & [cm/s$^{2}$] & Stellar surface gravity\\
  5     & e\_Logg                             & [cm/s$^{2}$] & Uncertainty in stellar surface gravity\\
  6     & [Fe/H]                              & dex          & Metallicity\\
  7     & e\_[Fe/H]                           & dex          & Uncertainty in metallicity\\
  8     & Vt                                  & km\,s$^{-1}$ & Microturbulence velocity\\
  9     & e\_Vt                               & km\,s$^{-1}$ & Uncertainty in microturbulence velocity\\
  10    & \(\mathrm{[Sr/H]}_{\mathit{NLTE}}\) & dex          & Strontium abundance\\
  11    & e\_[Sr/H]                           & dex          & Uncertainty in strontium abundance\\
  12    & \(\mathrm{[Y/H]}_{\mathit{NLTE}}\)  & dex          & Yttrium abundance\\
  13    & e\_[Y/H]                            & dex          & Uncertainty in yttrium abundance\\
  14    & [Zr\,{\sc i}/H]                     & dex          & Zirconium abundance\\
  15    & e\_[Zr\,{\sc i}/H]                  & dex          & Uncertainty in zirconium abundance\\
  16    & [Zr\,{\sc ii}/H]                    & dex          & Zirconium abundance\\
  17    & e\_[Zr\,{\sc ii}/H]                 & dex          & Uncertainty in zirconium abundance\\
  18    & \(\mathrm{[Ba/H]}_{\mathit{NLTE}}\) & dex          & Barium abundance\\
  19    & e\_[Ba/H]                           & dex          & Uncertainty in barium  abundance\\ 
  20    & [La/H]                              & dex          & Lanthanum abundance\\
  21    & e\_[La/H]                           & dex          & Uncertainty in lanthanum abundance\\
  22    & [Ce/H]                              & dex          & Cerium abundance\\
  23    & e\_[Ce/H]                           & dex          & Uncertainty in cerium abundance\\
  24    & [Pr/H]                              & dex          & Praseodymium abundance\\
  25    & e\_[Pr/H]                           & dex          & Uncertainty in praseodymium abundance\\
  26    & [Nd/H]                              & dex          & Neodymium abundance\\
  27    & e\_[Nd/H]                           & dex          & Uncertainty in neodymium abundance\\
  28    & \(\mathrm{[Eu/H]}_{\mathit{NLTE}}\) & dex          & Europium abundance\\
  29    & e\_[Eu/H]                           & dex          & Uncertainty in europium abundance\\
  \noalign{\smallskip}
  \hline
 \end{longtable}
 \tablefoot{Symbols \,{\sc i} and \,{\sc ii} denote the neutral and ionised state of elements respectively.}
 \centering

 \begin{table*}[h]
  \caption{Planetary characteristics of eleven additional stars. The remaining sample was published in \citetalias{Sharma24}.}
  \label{table:exoplanets}
  \centering
  \renewcommand{\tabcolsep}{1.85mm}
  \renewcommand{\arraystretch}{1.21}
  \begin{tabular}{l l c c c c c c c} 
  \hline
  \hline
  {\raisebox{-1.5ex}[0cm][0cm]{Host TYC ID}}  &  { \raisebox{-1.5ex}[0cm][0cm]{Planet}}  &  Planet Mass   &  Orbital Period  &  Semi Major Axis &  Ref.  \\
  &&  ($M_\mathrm{Jupiter}$)  &  (days)  &  (au)  &  \\
  \hline
  3131-1036-1   & HD 175370 b  & 4.60   & 349.5  & 0.98    &  [1]  \\
  3500-1780-1   & HD 153557 Ab & 0.064  & 7.3    & 0.068   &  [2]  \\
  3500-1780-1   & HD 153557 Ac & 0.055  & 15.3   & 0.111   &  [2]  \\  
  3568-2325-1   & HD 184960 b  & 0.0384 & 3.4982 & 0.04853 &  [3]  \\  
  4191-2696-1     & HD 154391 b  & 9.10   & 5163.0 & 7.46    &  [4]  \\  
  4405-1859-1   & HD 118904 Ab & 3.10   & 676.7  & 1.70    &  [5]  \\
  4425-567-1    & HD 150010 Ab & 2.40   & 562.0  & 1.40    &  [6]  \\
  4434-2189-1   & HD 174205 b  & 4.20   & 582.0  & 1.70    &  [6]  \\
  4436-1423-1   & HD 161178 b  & 0.57   & 279.3  & 0.84    &  [7]  \\
  4464-1666-1   & HD 202432 b  & 1.90   & 418.8  & 1.20    &  [5]  \\ 
  4534-1837-1   & HD 46588 b   & 0.25   & 223.0  &  --     &  [8]  \\
  4573-1915-1   & HD 164428 b  & 5.70   & 599.6  & 1.60    &  [5]  \\
  \hline
  \end{tabular}
  \flushleft{\bf{References.}} [1] - \citet{Hrudkova17}. [2] - \citet{Feng22}. [3] - \citet{Barnes23}. [4] - \citet{Xiao24}. [5] - \citet{Jeong18}. [6] - \citet{Jeong22}. [7] - \citet{Teng22}. [8] - \citet{Subjak23}.
 \end{table*}

\end{document}